\documentclass[aps,pre,superscriptaddress,twocolumn,balancelastpage]{revtex4-2}

\usepackage[colorlinks,bookmarks=false,citecolor=blue,linkcolor=blue,urlcolor=blue]{hyperref}
\usepackage[all]{hypcap}   

\usepackage{physics}

\usepackage{amsmath,amssymb}
\usepackage{graphicx}
\usepackage{bbm}

\graphicspath{{figures/}{/}}

\usepackage{verbatim}
\usepackage{color}

\usepackage{placeins}  
\usepackage{flafter}     
\usepackage[T1]{fontenc}
\usepackage{lmodern}

\usepackage{color}

\newcommand{\HIDDEN}[1]{}

\newcommand{\ue}{\text{e}}
\newcommand{\ui}{\text{i}}
\newcommand{\ud}{\text{d}}

\newcommand{\uA}{{\text{A}}}
\newcommand{\uB}{{\text{B}}}
\newcommand{\calU}{\mathcal{U}}
\newcommand{\calH}{\mathcal{H}}
\newcommand{\rmt}{\text{RMT}}

\begin{document}

\title{Entanglement in bipartite systems with symmetry: \\
coupled chaotic kicked Bose-Hubbard systems}

\author{Jan Himmelsbach}
\affiliation{TU Dresden,
Institute of Theoretical Physics and Center for Dynamics,
 01062 Dresden, Germany}

\author{Maximilian F.~I. Kieler}
\affiliation{TU Dresden, Institute of Theoretical Physics and Center for
Dynamics, 01062 Dresden, Germany}
\affiliation{CESAM research unit, University of Liège, B-4000 Liège, Belgium}

\author{Arnd B\"acker}
\affiliation{TU Dresden,
Institute of Theoretical Physics and Center for Dynamics,
 01062 Dresden, Germany}

\date{\today}

\begin{abstract}

We investigate the average eigenstate entanglement in a bipartite many-body
system which exhibits a breaking of two local conservation laws into a global
conserved quantity. Such a setting is realized by the particle number
conservation in Bose-Hubbard systems. We devise a corresponding random matrix
model which captures the universal features of this symmetry breaking and
allows for applying powerful random matrix methods. By combining the concept
of symmetry resolved entanglement with perturbation theory for quantum chaotic
systems we obtain a universal entanglement transition depending on a single
fundamental parameter. Furthermore, the symmetry resolved  entanglement allows
for separating the genuine entanglement from the part which originates from
the conserved quantity. This latter contribution is quantified by the number
entropy. For this it is shown that the symmetry breaking generates a
localization of the eigenstates due to a banded structure of the
time-evolution operator. By extrapolating the results beyond the perturbative
regime we obtain an analytic description of the full transition.

\end{abstract}

\maketitle

\section{Introduction}

A central challenge of contemporary many-body physics is to classify the
dynamical features of many-body systems. This is of particular importance for
gaining a deeper understanding of the processes of dynamical and eigenstate
thermalization and the emergence of quantum chaos.
Entanglement is one of the key indicators of dynamical phases in many-body
systems \cite{AmiFazOstVed2008, NanHus2015, AbaPap2017, AbaAltBloSer2019}. It
is extensively studied in various system classes ranging from spin chains
and random matrix circuits to Bose and Fermi-Hubbard systems.
The notion of entanglement of pure states in systems without a symmetry is
well established by means of the entanglement entropies and the Schmidt
decomposition. In contrast the characterization of entanglement in systems
with a symmetry allows for a finer resolution by means of symmetry resolved
entanglement entropies \cite{GolSel2018}. They provide a separation of the
entanglement between subsystems in a fixed symmetry sector and the
correlations stemming from the superposition of multiple symmetry sectors. It
is commonly applied in the realm of (conformal) field theories
\cite{GolSel2018, FelGol2019, BonRugCal2019, CapRugCal2020, MurDubCal2024} as
well as for the characterization of many-body localization \cite{KieUnaFleSir2020,KieUnaFleSir2021}.
It is also used to give a bound on the entanglement
\cite{KieUnaSirFle2020} in fermionic systems. Moreover,
it can be experimentally employed for the quantification
of entanglement \cite{LukRisSchTaiKauChoKheLeoGre2019,
NevCarVitKokElbDalCalZolVerKueKra2021, VitElbKueNevCarKraZolCalVerDal2022}.

\begin{figure}\label{fig:bose_hubbard_scheme}
	\includegraphics{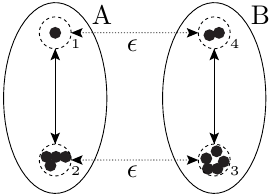}
	\caption{%
	Schematic illustration of a bipartite system with a conserved quantity.
	The Bose-Hubbard system consists of four sites indicated by the four
	dash-lined circles. Particles are depicted by full black circles. The
	system is divided into two subsystems A and B with two sites each. The
	parameter $\epsilon$ controls the interaction between the subsystems.
	Another interaction allows for particle exchange within the subsystem.}
\end{figure}

In this paper we investigate the generation of symmetry resolved entanglement
in a bipartite setting. The considered system consists of two subsystems,
where each subsystem is assumed to be quantum chaotic, but has a single
conserved quantity. By introducing a coupling between the subsystems, these two
conservation laws reduce to a single conserved quantity of the full system
only. This process appears for example in (Bose)-Hubbard systems, where the
conserved quantity is realized by the particle number. For concreteness, one
can consider two separate Bose-Hubbard systems each with fixed particle
numbers, see for example Fig.~\ref{fig:bose_hubbard_scheme}. By letting
them interact, they can interchange particles and hence only the total
particle number remains preserved.
From the entanglement point of view, one can identify quantum correlations
between the subsystems responsible for the entanglement. Moreover there are
additional correlations due to the conserved quantities.
To illustrate this explicitly, we consider an eigenstate of the Bose-Hubbard
system with $N$ particles, for which a measurement of the particle number
$N_\uA$ in subsystem $\uA$ directly implies the particle number $N_\uB = N -
N_\uA$ in subsystem $\uB$. This kind of correlation is not resolved by the
usual bipartite entanglement entropy. In contrast, the symmetry resolved
entanglement allows to disambiguate both aspects, namely the part which is
genuinely attributed to the quantum correlations between both subsystems and
the part describing the correlations stemming from the conserved quantities.
Hence, using the symmetry resolved entanglement entropy allows for a  deeper
understanding of the effects arising from the breaking of two conserved
quantities into a single conserved quantity, which is accompanied by a
localization of the eigenstates.
To obtain a universal random matrix description, we identify the chaotic
subsystems with random matrices and hence obtain a structured random matrix
model. We show by means of a perturbation theory for bipartite systems
\cite{SriTomLakKetBae2016, TomLakSriBae2018}, that this random matrix model
possess a universal entanglement transition: There exists a single transition
parameter which combines dimensional aspects of the structure arising from the
conservation laws with the interaction strength. The entanglement transition
from the uncoupled systems towards the strongly coupled system depends only on
this parameter, hence is universal, independently from the specific dimensions
of the underlying systems.

The paper is organized as follows: In Sec.~\ref{sec:bipartite_setting} we
introduce the structure of the Hilbert space and the time evolution
operator imposed by a conserved quantity and devise a random matrix model
which intrinsically implements this structure. Furthermore, we introduce the
symmetry resolved entanglement formalism. In
Sec.~\ref{sec:entanglement_transition}, we start by discussing
phenomenologically the features of the entanglement transition for varying
coupling strength and afterwards derive a perturbation theory for the symmetry
resolved entanglement. Subsequently, we discuss a non-perturbative
extrapolation allowing for a characterization of the full transition.
Eventually, we apply our results to a Bose-Hubbard system, in
Sec.~\ref{sec:Bose_Hubbard_model}. Section~\ref{sec:summary} gives
a summary and outlook.

\section{Bipartite setting}\label{sec:bipartite_setting}

We start by shortly recapitulating the results and formulations of explicitly
bipartite (Floquet) systems in the absence of any conservation law introduced
in \cite{SriTomLakKetBae2016, LakSriKetBaeTom2016}. We call a quantum system
which is composed of two subsystems A and B subject to a tunable coupling
\textit{bipartite}. Explicitly, we consider unitary Floquet systems
characterized by a (discrete) time evolution operator $\calU$ acting on a
tensor product space $\calH^\uA \otimes \calH^\uB$ with $\calH^\uA$ and
$\calH^\uB$ of dimension $M_\uA$ and $M_\uB$, respectively. The dynamics on
this Hilbert space is given in the form of a time evolution operator
\begin{equation}
	\mathcal{U}(\epsilon) = \mathcal{U}_{\uA \uB}(\epsilon) \mathcal{U}_0,
	\label{eq:time_evol_w_interaction}
\end{equation}
where $\mathcal{U}_0 = \mathcal{U}_0^\uA \otimes \mathcal{U}_0^\uB$ is the
uncoupled time evolution operator of subsystems A and B. The operator
$\mathcal{U}_{\text{AB}}(\epsilon)$ provides a tunable interaction between the
subsystems, which is assumed to vanish for $\epsilon = 0$,
i.e.,~$\mathcal{U}_{\text{AB}}(0) = \mathbbm{1}$ and establishes a
strong interaction for $\epsilon \gg 1$. The particular
structure \eqref{eq:time_evol_w_interaction} is motivated by their generic
appearance in physical systems like quantum maps \cite{BerBalTabVor1979} or
quantum circuits \cite{ChaLucCha2018a} and spin chains \cite{Pro2000}.

Systems of this structure are commonly studied in the realm of quantum chaos
\cite{SriTomLakKetBae2016, TomLakSriBae2018, PulLakSriBaeTom2020,
HerKieFriBae2020, PulLakSriKieBaeTom2023, FriKie2024} and many-body
thermalization/ergodicity \cite{ChaLucCha2018a, YosGarCha2025}, and recently
using field theoretic methods \cite{AltTelMic2024, AltKimMic2025:p}. One
assumes $\mathcal{U}^\uA_0, \mathcal{U}^\uB_0$ to be quantum chaotic, i.e., their spectra follow
the predictions of random matrix theory. Consequently, for strong coupling it
is expected that the subsystem structure becomes invisible and the system
behaves as a single random matrix of dimension $M_\uA M_\uB$.
This increasing loss of structure from uncoupled systems towards a strongly
interacting system arises in form of a universal transition depending on a
single parameter $\Lambda(M_\uA, M_\uB, \epsilon)$, which combines the system
dimension and the actual coupling parameter \cite{SriTomLakKetBae2016,
TomLakSriBae2018, PulLakSriBaeTom2020, HerKieFriBae2020,
PulLakSriKieBaeTom2023, FriKie2024}.
This transition is universal in the sense that it is independent from the
underlying dynamics and in fact can be fully characterized by replacing
$\mathcal{U}^\uA_0, \mathcal{U}^\uB_0$ as well as $\mathcal{U}_{\uA\uB}(\epsilon)$ by random unitary matrices. This kind of random
matrix universality was originally shown using perturbation theory arguments
\cite{SriTomLakKetBae2016, LakSriKetBaeTom2016, TomLakSriBae2018} and can also
be found via Weingarten calculus \cite{FriKie2024} and semiclassical periodic
orbit theory \cite{KieFriBae2026:p}.

\subsection{Conserved quantity}

In the following, we extend the previous bipartite model to exhibit a
transition from two conserved quantities within the subsystems into a single
global conserved quantity, only. As outlined in the introduction, such a
transition is of great interest for (Bose)-Hubbard systems, where it is
realized in form of the particle number $N = N_\uA + N_\uB$. In order to give
a universal and system-independent description, we introduce in the following
an abstract formulation of this setting.

Consider a conserved quantity $Q$ for some time-evolution operator
$\mathcal{W}$, i.e.,
\begin{align}
 [\mathcal{W}, Q] &= 0. \label{eq:glob_integral_of_motion}
\end{align}
The Hilbert space $\mathcal{H}$ for such a system decomposes into invariant
subspaces $\mathcal{H}_q$, i.e.,~the eigenspaces of $Q$ corresponding to the
(real) eigenvalues $q$. We always assume countable decompositions and we
identify $q$ with some number, i.e.,~$q = 0, 1, 2, \ldots$, for simplicity of
notation. Consequentially, the Hilbert space decomposes as
\begin{align}
 \mathcal{H} &\simeq \bigoplus \limits_{q} \mathcal{H}_q.
\end{align}
We denote states from this Hilbert space by $\ket{i; q} \in \mathcal{H}$,
with $i=1, \ldots, k_q$ enumerating the basis of $\mathcal{H}_q$. We assume
that these Hilbert spaces are finite with dimension $k_q = \dim \mathcal{H}_q$.
Due to \eqref{eq:glob_integral_of_motion}, the time-evolution operator leaves
the subspaces invariant and hence decomposes itself into
\begin{align}
	\mathcal{W} &= \sum \limits_{q} \mathcal{W}^{(q)},
\end{align}
where $\mathcal{W}^{(q)}$ acts on $\mathcal{H}_q$. For the remainder we fix
some eigenspace $\mathcal{H}_q$ and the corresponding time evolution operator
$\mathcal{W}^{(q)}$.
We consider now a system built from two such systems (denoted A and B) each
possessing an individual conserved quantity $Q_\uA$ and $Q_\uB$, respectively.
Moreover, we define the global conserved quantity as their additive combination
\begin{align}
	Q &= Q_\uA \otimes \mathbbm{1}_\uB +  \mathbbm{1}_\uA \otimes Q_\uB,
\end{align}
which is clearly another but not independent conserved quantity.
By construction the bipartite Hilbert space of these two systems restricted to
a fixed sector corresponding to value $q$ of $Q$ is given by
\begin{align}
	\mathcal{H}_q &\simeq \bigoplus_{q_\uA}
	\mathcal{H}_\uA^{(q_\uA)} \otimes \mathcal{H}_\uB^{(q - q_\uA)}.
			\label{eq:bip_hilbert_space_w_cons}
\end{align}
In contrast to the formerly introduced bipartite systems, we encounter here an
additional decomposition into invariant subspaces. We will call this Hilbert
space in the following the Fock space. The corresponding time evolution
operator of the non-interacting system attains by conservation of
$Q_\uA$ and $Q_\uB$ the corresponding block structure
\begin{align}
	\mathcal{U}_0^{(q)} &= \sum_{q_\uA}
				\calU^{(q_\uA)}_\uA \otimes \calU^{(q - q_\uA)}_\uB.
\end{align}
Interactions between the subsystems can either preserve the subsystems
conserved quantities, break both conservation laws or reduce them into a
single conserved quantity. The latter case is of particular interest as it
occurs in (Bose-)Hubbard systems, where the interaction exchanges particles
between the subsystems but leaves the total particle number constant.
Therefore, we assume for the full time evolution operator $\mathcal{U} =
\mathcal{U}_{\uA \uB} \mathcal{U}_0$
\begin{align}
	[\mathcal{U}, Q] &= 0 \\
	[\mathcal{U}, Q_\uA \otimes \mathbbm{1}_\uB] &\neq
		0 \neq [\mathcal{U}, \mathbbm{1}_\uA \otimes Q_\uB].
\end{align}

\subsection{Dynamics}\label{subsec:dynamcis}

We are aiming to identify the universal features of entanglement in systems in
which a conservation law is broken by an interaction. Therefore we want to
eliminate all system specific features of real physical systems but preserve
the universal algebraic structure arising from this process. If the subsystems
consists of many degrees of freedom, or by an appropriate choice of the
systems parameters, the subsystem dynamics is effectively quantum chaotic.
Therefore, we assume that the physical subsystem operators of a system are
well approximated by random matrices, i.e.,
\begin{align}
	\calU_\uA^{(q_\uA)} \to \calU_{\uA, \rmt}^{(q_\uA)} \\
	\calU_\uB^{(q_\uB)} \to \calU_{\uB, \rmt}^{(q_\uB)},
\end{align}
where $\calU_{j, \rmt}^{(q_j)}$ are independent in both the subsystems $j \in
\{\uA, \uB \}$ and over the subblock configurations $q_\uA$ and $q_\uB$.
Depending on the time-reversal invariance of the physical systems,
we either draw the matrices from the circular unitary ensemble (CUE) or the
circular orthogonal ensemble (COE). Thereby, we eliminate non-universal
features tied to the specific physical dynamics.

The coupling
\begin{align}
	\mathcal{U}_{\uA \uB}^{(q)} &= \exp{\ui V}
	\label{eq:coupling-U}
\end{align}
is supposed to establish the interaction between the subsystems and moreover
breaks the subsystems conserved quantities $Q_\uA$ and $Q_\uB$ by generating
couplings between different blocks $q_\uA$ and $q_\uA'$. We refer to the first
as the \textit{subsystem interaction} and the second as \textit{Fock space
interaction}. There is a variety of possible implementations for such a
coupling \eqref{eq:coupling-U}, but here we focus on an interaction which
couples nearest-neighboring blocks with a generically random subsystem
interaction. Such a coupling is given by the potential
\begin{align}
	\matrixel{i'j'; q_\uA'}{V^{(q)}}{ij; q_\uA} =
		\begin{cases}
		      h^{(q_\uA)}_{i'j',ij} & q_\uA = q_\uA' + 1 \\
		      (h^{(q_\uA+1)}_{i'j',ij})^\dagger & q_\uA = q_\uA' - 1 \\
		      0 & \text{else}
		\end{cases}
		\label{eq:coupling_potential}
\end{align}
where $h^{(q_\uA)}$ are independent, rectangular Gaussian matrices,
i.e.,~$h^{(q_\uA)}_{i'j',ij}$ are drawn independently from the normal
distribution with zero mean and a variance
\begin{align}
 \sigma_{q_\uA}^2 &= \frac{\epsilon^2 \tilde{\sigma}_{q_\uA}^2}{
	\beta k_{q_\uA} k_{q_\uA-1} }. \label{eq:coupling_variance}
\end{align}
For subsystem dynamics following the COE, we choose $h^{(q_\uA)}_{i'j',ij}$ to
be real and set $\beta = 1$. In contrast for the CUE,
the $h^{(q_\uA)}_{i'j',ij}$ are chosen to be complex and hence $\beta = 2$.
We distinguish an overall coupling strength $\epsilon$, which is assigned
to all blocks and hence allows to establish $\mathcal{U}_{\uA\uB}(0) =
\mathrm{1}$. Furthermore, we weight the
variance by the dimension of the matrix to eliminate intrinsic
block-size effects. Note that this scaling can be effectively reverted or
enhanced by choosing the block-specific variances $\tilde{\sigma}_{q_\uA}^2$
accordingly. This allows to give each block a specific artificial dimension, as
discussed in Sec.~\ref{sec:entanglement_transition}.

By above choices the dynamics of the physical system is fully reduced
to a random matrix setting, which is expected to show  (on average) the same
characteristics. Therefore we define the random matrix ensemble as the Haar
average over the unitary matrices $\calU_{j, \rmt}^{(q_j)}$ and a Gaussian
average over the coupling, see Eq.~\eqref{eq:coupling_potential}.
Furthermore, it is beneficial for perturbation theory to transform the
ensemble into a form in which $\mathcal{U}_0^{(q)}$ is diagonal. Therefore, we
assume the eigendecomposition $\mathcal{U}_0^{(q)} = \mathcal{V}^\dagger E
\mathcal{V}$ with $E$ being the diagonal matrix of all eigenvalues. The unitary
transformation
\begin{align}
	\mathcal{U}^{(q)} \to \mathcal{V} \mathcal{U}^{(q)} \mathcal{V}^\dagger
		= E \mathcal{V} \mathcal{U}_{\uA\uB}^{(q)} \mathcal{V}^\dagger,
\end{align}
together with the unitary invariance of the Gaussian coupling establishes
finally the \textit{conserved quantity transition ensemble} (CQTE)
\begin{align}
	\text{CQTE} = \{ E \ue^{\ui V} \}, \label{eq:cqte_ensemble}
\end{align}
where $E$ is distributed according to the eigenvalue distribution of the COE
or CUE.

For the generation of numerical data, we consider two choices of dimensions.
First we consider the simplest case of a system of $q$ blocks ($q_\uA = 1,\ldots,q$), all with equal
block dimension $k_{q_\uA}^\uA = k_{q_\uB}^\uB \equiv k$. Due to the finite number of blocks, we set $h^{(1)} = h^{(q+1)} =
0$. Elsewhere, we fix
the block variances to be uniformly $\tilde{\sigma}_{q_\uA}^2 = 1$. As a
second choice, we consider systems with dimensions selected according to a
four site Bose-Hubbard system of $N$ particles partitioned into subsystems of
two sites each, see Fig.~\ref{fig:bose_hubbard_scheme}. This gives a total
number of blocks $q = N + 1$ and we have the block dimension $k_{q_\uA}^\uA =
q_\uA + 1$ and $k_{q - q_\uA}^\uB = q - q_\uA + 1$ for $q_\uA = 0, \ldots, N$.
We assume for this structure also $\tilde{\sigma}_{q_\uA}^2 = 1$, with  $h^{(0)} = h^{(q+1)} =
0$ at the boundaries.

\begin{figure}\label{fig:time_evolution_operator}
	\includegraphics{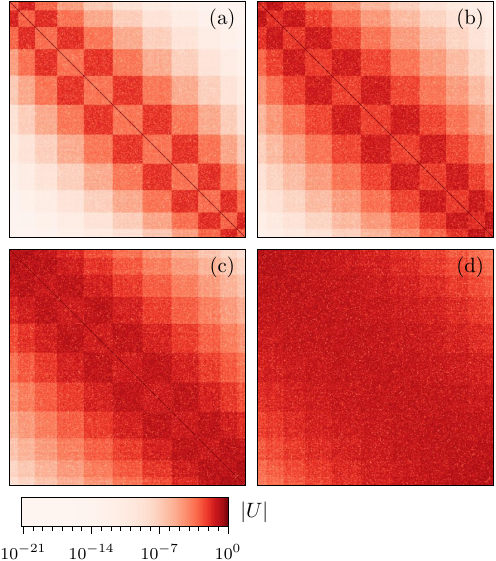}
	\caption{Realizations of the time evolution operator of the CQTE
		\eqref{eq:cqte_ensemble} for the four site Bose-Hubbard system for
		$N=10$ particles and COE statistics. Logarithmic color scale and
		subsystem diagonalized basis for (a-d) $\sqrt{\Lambda_q} = 0.1,~0.4,~
		1.2,~4.0$, respectively. The transition parameter $\sqrt{\Lambda_q}$
		is defined in Eq.~\eqref{transition_parameter}.
        }
\end{figure}

In Fig.~\ref{fig:time_evolution_operator} we present some realizations of this
ensemble in logarithmic scale for various values of $\epsilon$ and dimensions
according to the four site Bose Hubbard system. In absence of any coupling,
the time-evolution operator is diagonal (not shown). The interaction
introduces a broadening, turning the matrix into a coarse grained banded
matrix with an apparent block structure. Importantly, due to the construction
of the interaction there is only a coupling of neighboring blocks. Thus the
block-wise off-diagonals of the matrix decrease effectively in orders of
$\epsilon$. This can be seen by expanding $\ue^{\ui V} = 1 + \ui V - V^2/2 +
\ldots$. The first order $V \sim \epsilon$ generates couplings between
neighboring blocks hence give contributions on the first off-diagonal blocks.
The
second order $V^2 \sim \epsilon^2$ couples to next-nearest blocks, and via the
nearest block back to itself, hence the second block-wise off-diagonals are
$\sim \epsilon^2$. As a consequence one gets an exponential decrease of
off-diagonal contributions in orders of $\epsilon$.
This can be seen in Fig.~\ref{fig:time_evolution_operator}(a-c),
where we plot the logarithmic absolute value of $U$. Moreover, for large
coupling strength the underlying
block structure dissolves into a unstructured random matrix,
see Fig.~\ref{fig:time_evolution_operator}(d).
The prominent diagonal of the uncoupled system gradually dissolves as well,
indicating the vanishing relevance of the subsystem dynamics.

Due to its central diagonal of the uncoupled matrix $\mathcal{U}_0$ and the
banded structure, this ensemble combines structural features typical for
Rosenzweig-Porter type ensembles \cite{RosPor1960, KraKhaCueAmi2015,
BuiBar2022} with banded-matrix ensembles \cite{Efe1983, FyoMir1994,
AltZir1996}. We therefore expect that the eigenstate properties follow a
non-trivial random matrix prediction beyond the bipartite case.

\subsection{Entanglement: Moments and entropies}

The quantification of entanglement in bipartite systems is well established
by means of entanglement entropies. We shortly recapitulate the standard
procedure based on the evaluation of Schmidt values and afterwards discuss
necessary modifications for the application to systems with two conserved
quantities decaying into a single conserved quantity given by
symmetry-resolved entanglement entropies \cite{GolSel2018}.

A vector $\ket{\psi}$ in a bipartite Hilbert space $\mathcal{H}^\uA
\otimes \mathcal{H}^\uB$ with $\dim \calH^\uA = M_\uA, \dim \calH^\uB = M_\uB$
$(M_\uA \leq M_\uB)$, can always be represented via Schmidt-decomposition as
\begin{equation}
	\ket{\psi} = \sum_{j=1}^{M_\uA}
			\sqrt{\lambda_j} \ket{\psi_j^\uA}\ket{\psi_j^\uB},
			\label{eq:schmidt_decomposition}
\end{equation}
where $\{\ket{\psi_j^\uA}\}$ and $\{\ket{\psi_j^\uB}\}$ are orthogonal bases
in subsystem $\uA$ and $\uB$, respectively. The non-negative real numbers
$\lambda_j$ are called the Schmidt values, which we assume to be ordered
$1 \geq \lambda_1 \geq \lambda_2 \geq \cdots \geq \lambda_{M_\uA} \geq 0$. They
satisfy the normalization condition
\begin{equation}
	\sum_{j=1}^{M_\uA} \lambda_j = 1.
\end{equation}
If $\lambda_1 = 1$ and all other Schmidt values are equal to zero,
i.e.,~$\ket{\psi}$ is a product state, it is called separable or unentangled.
If two or more Schmidt values are different from zero the state is said to be
entangled. The Schmidt states $\ket{\psi_j^\uA}$ and $\ket{\psi_j^\uB}$ are the
eigenstates of the reduced density matrices $\rho_\uA$ and $\rho_\uB$,
\begin{align}
	\rho_\uA &= \text{tr}_\uB(\dyad{\psi}) \label{eq:red_dens_op}\\
	\rho_\uB &= \text{tr}_\uA(\dyad{\psi}),
\end{align}
which are obtained by tracing subsystem $\uB$ and $\uA$, respectively.
One commonly used measure to quantify the entanglement of the state
$\ket{\psi}$ is the linear entanglement entropy $S_2$ defined as
\begin{equation}
	S_2 = 1 - \text{tr}(\rho_\uA^2) = 1 - \sum_{j=1}^{M_\uA} \lambda_j^2.
	\label{eq:lin_ent_entr}
\end{equation}
Note that $S_2 = 0$ for unentangled states. The average eigenstate
entanglement of a time-evolution operator $\calU$ is defined as the average of
$S_2$ over all eigenstates $\ket{\Psi_n}$ of $\calU$,~i.e.,~satisfying
$\calU \ket{\Psi_n} = \ue^{\ui \phi_n} \ket{\Psi_n}$.

\begin{figure}
	\includegraphics{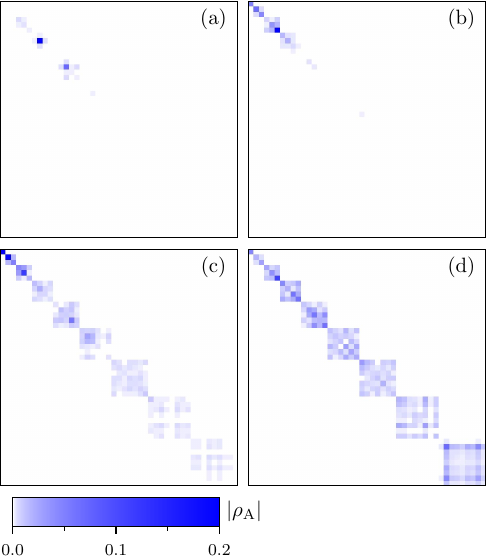}
	\caption{Density operators $\rho_A$ of randomly chosen eigenvectors of a
		realization of the CQTE with COE statistics. The density matrix is
given in the
		eigenbasis of the uncoupled system $\mathcal{U}(0)$.
		The dimensions are selected according to a
		four site Bose-Hubbard system with $N=8$ particles. (a-d) correspond
		to coupling strengths $\sqrt{\Lambda_q} = 0.1,~0.4,~ 1.2,~4.0$,
		respectively. The transition parameter is defined in Eq.~\eqref{transition_parameter}.
		}
	\label{fig:density_matrix_transition}
\end{figure}

In the presence of a conserved quantity the lack of a simple tensor product
form in \eqref{eq:bip_hilbert_space_w_cons} complicates a formulation of the
Schmidt decomposition. In contrast the approach based on the reduced
density matrix still applies and an entropy can be computed from the
eigenvalues of $\rho_\uA$, hence giving the entanglement entropy $S_2$.
However, this
formulation of an entanglement entropy has a significant drawback that it not
clearly separates between correlations which come from the conservation law
(i.e.,~the block matrix structure) and correlations between the
subsystems (which generate actual entanglement). This becomes intuitively
visible by plotting a reduced density matrix for strong coupling, see
Fig.~\ref{fig:density_matrix_transition}(d).
The key observation is that under the partial trace the block decomposition
into $q_\uA$-subsectors survives, even in the presence of interactions.

To explicitly show this specific form, consider
the reduced density matrix of a state $\ket{\psi} \in \calH^{(q)}$ in the
eigenbasis of $Q_\uA$ and $Q_\uB$,
\begin{align}
	&\matrixel{i'; q_A'}{\rho_A}{i; q_A} = \nonumber \\
	&\qquad \sum_{j, q_B}
		\matrixel{i'j; q_A'}{\rho_{\Psi}}{ij; q_A}
			\delta_{q_\uB, q - q_\uA'} \delta_{q_\uB, q - q_\uA}.
			\label{eq:block_reduced_density_matrix}
\end{align}
The global conserved quantity $Q$ forces by
\begin{align}
	q &= q_\uA + q_\uB = q_\uA' + q_\uB,
\end{align}
the strong restriction $q_\uA = q_\uA'$. In consequence the reduced
density matrix decomposes into subblocks of dimension
$\dim \mathcal{H}_\uA^{(q_\uA)}$.

\begin{figure}
	\includegraphics{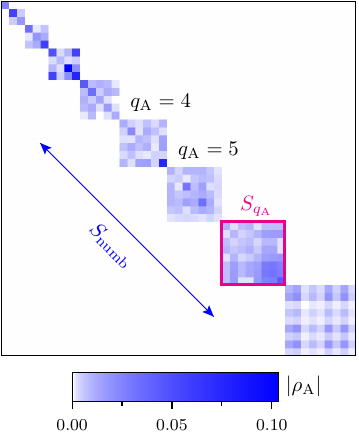}
	\caption{
		Density operator $\rho_A$ of a randomly chosen eigenvector of a
		realization of the CQTE with COE statistics. The density matrix is
given in the
		eigenbasis of the uncoupled system $\mathcal{U}(0)$.
		The dimensions are selected according to a
		four site Bose-Hubbard system with $N=8$ particles. The coupling
		strength is $\sqrt{\Lambda_q} = 4.0$, see Eq.~\eqref{transition_parameter}.
		The number entropy measures the localization along $q_\uA$, whereas
		$S_{q_\uA}$ measures entanglement in one block.}
	\label{fig:density_matrix_entropies}
\end{figure}

The $q_\uA$-reduced density matrix corresponding to such a block is given by
$\tilde{\rho}_{q_\uA} = \Pi_{q_A}\rho_A\Pi_{q_A}$, where we define the
projection operator $\Pi_{q_A}$ as
\begin{equation}
	\Pi_{q_A} \ket{ij; q_A'} = \delta_{q_A', q_A}\ket{ij; q_A'}.
\end{equation}
We emphasize that $\tilde{\rho}_\uA$ has all properties of a reduced density
matrix but is not normalized as only the total reduced density matrix is
normalized
\begin{align}
	\sum_{q_\uA} \tr \tilde{\rho}_{q_\uA} &= 1.
\end{align}
In order to turn these subblocks into reduced density matrices themselves, one
introduces block amplitudes
\begin{align}
	p_{q_A} &= \text{tr}( \tilde{\rho}_{q_\uA}),
\end{align}
carrying the normalization and defines
\begin{align}
	\rho_{q_\uA} &= \frac{\tilde{\rho}_{q_\uA}}{p_{q_\uA}},
\end{align}
for $p_{q_\uA} \neq 0$, otherwise one sets $\rho_{q_\uA} = 0$. Again due to
normalization, we have $1 = \sum_{q_\uA} p_{q_\uA}$, i.e.,~the
set of $p_{q_\uA}$ are probabilities and we can define their entropy, commonly
called particle-number resolved entropy \cite{GolSel2018}, or short number
entropy, as
\begin{align}
	S_{\text{numb}} &= 1 - \sum_{q_\uA} p_{q_\uA}^2.
\end{align}
This entropy has an appealing interpretation as a localization measure along
the broken symmetry degree: If a state is completely localized in some block,
i.e.,~$p_{q_\uA} = 1$ for some $q_\uA$, then $S_{\text{numb}} = 0$. In
contrast, the number entropy is maximal for fully delocalized states with all
$p_{q_\uA}$ being equal. Thus $S_\text{numb}$
characterizes the localizing behavior along $Q_\uA$. However, it is
fully insensitive for spatial correlations between the subsystems $\uA$ and
$\uB$. Therefore, we employ the normalized $q_\uA$-reduced density matrices by
defining the entropy of a block by
\begin{equation}
	S_{q_A} = 1 - \text{tr}(\rho_{q_A}^2). \label{eq:entropy_of_a_block}
\end{equation}
The entropies of blocks are not particular meaningful whenever the block
amplitude $p_{q_\uA}$ is small. Therefore it is convenient to introduce a
configurational entropy (or accessible entanglement) \cite{BarHerDel2018,
BarCasDel2019, MurCalPir2022}
\begin{align}
 S_\text{conf} &= \sum_{q_\uA} p_{q_\uA}^2 S_{q_\uA},
		\label{eq:configurational_entropy}
\end{align}
where we set $p_{q_\uA}^2 S_{q_\uA} = 0$ whenever $p_{q_\uA} = 0$. It gives a
weighed average over all entropies of blocks. A schematic illustration of
$S_\text{numb}$ and $S_{q_\uA}$ can be seen in
Fig.~\ref{fig:density_matrix_entropies}.

Both the number entropy and the configurational entropy contribute additively
to the total linear entanglement entropy $S_2$ as
\begin{align}
	S_2 &= 1 - \text{tr}(\rho_\uA^2) \\
	    &= S_{\text{numb}} + S_\text{conf}.
	    \label{eq:entropy_composition}
\end{align}

For strongly coupled members of the CQTE, i.e.,~large $\epsilon$, we expect
that the eigenstates behave as random states. The corresponding random
state predictions are obtained by forcing the block structure arising from the
symmetry onto unstructured random states. Let $k_{q_\uA}^\uA$ be the
Hilbert space dimension of $\mathcal{H}_\uA^{(q_\uA)}$ and $k_{q_\uA}^{\uB}$
the dimension of $\mathcal{H}_\uB^{(q-q_\uA)}$ in
Eq.~\eqref{eq:bip_hilbert_space_w_cons}. Then one has $k_{q_\uA} =
\text{dim}(\mathcal{H}_\uA^{q_\uA} \otimes \mathcal{H}_\uB^{(q-q_\uA)}) =
k_{q_\uA}^{\uA} k_{q_\uA}^{\uB}$ and the average entanglement entropy, number
entropy and configurational entropy are given
by~\cite{BiaDon2019}
\begin{align}
	S_\text{numb}^\infty  &= 1 -
	    \sum_{q_\uA} \frac{k_{q_\uA}(k_{q_\uA} + 3 - \beta)}{K(K + 3 - \beta)}
			\label{eq:rmt__S_numb} \\
	S_2^\infty &= 1 - \sum_{q_\uA}
		\frac{k_{q_\uA}(k_{q_\uA}^{\uA} + k_{q_\uA}^{\uB} + 2 - \beta)}
		{K(K + 3 - \beta)}
			\label{eq:rmt_S_2} \\
	S_\text{conf}^\infty &= \sum_{q_\uA}
	\frac{k_{q_\uA}(k_{q_\uA} - k_{q_\uA}^{\uA} - k_{q_\uA}^{\uB}  + 1)}
	{K(K + 3 - \beta)} ,
\end{align}
where $K = \sum_{q_\uA}k_{q_\uA}$ is the dimension of the Hilbert space
$\mathcal{H}_q$ and $\beta = 1$ for the COE and $\beta = 2$ for the CUE.
The expectation values in Eqs.~\eqref{eq:rmt__S_numb} and \eqref{eq:rmt_S_2}
are close to unity for large system sizes.

\section{Entanglement transition}\label{sec:entanglement_transition}

\subsection{Phenomenological discussion} \label{sec:pheno_discussion}

\begin{figure}
	\includegraphics{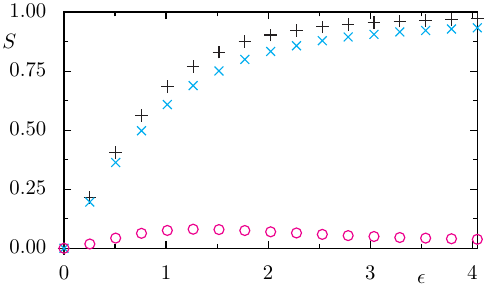}
	\caption{Transition of $S_2$ (black $+$), $\tilde{S}_\text{numb}$ (cyan
    $\times$) and $\tilde{S}_\text{c}$ (magenta $\circ$) as functions of
    the coupling strength $\epsilon$ for the CQTE with $k = k_{q_\uA}^\uA
    = k_{q_\uB}^\uB = 5$ and $q=50$ blocks and COE statistics.
	}
\label{fig:entanglement_transition_wo_resc}
\end{figure}

We start by discussing the phenomenological features of the entanglement
transition from uncoupled towards a strongly interacting system. The detailed
analytical derivation will be carried out afterwards.
We show in Fig.~\ref{fig:entanglement_transition_wo_resc} the transition of
the entropy $S_2$ decoded into number entropy $S_\text{numb}$ and
configurational entropy $S_\text{conf}$ components for various values of an
effective coupling
strength $\epsilon$ of the CQTE.
The number entropy is monotonously increasing. In contrast the configurational
entropy shows initially a nonlinear slope and reaches a maximum after
which it slightly decreases and then slowly settles to the random matrix value.
The total entropy is dominated by the number entropy with much smaller
contributions coming from the configurational entropy. In fact, for small
coupling strengths $\epsilon \ll
1$, the total entropy is essentially equal to the number entropy,
\begin{align}
	S_2 &\approx S_\text{numb}.
\end{align}

\begin{figure}
	\includegraphics{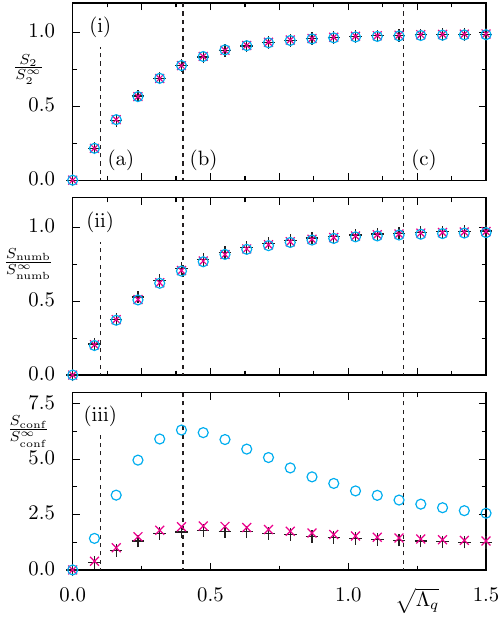}
	\caption{
        Transition of (i) the rescaled total entropy (ii) the rescaled number
        entropy and (iii) the rescaled configurational entropy for three
        different configurations. We show numerical data for the
        equidimensional CQTE with $q=15, k=10$ (magenta $\times$) and $q=50,
        k=5$ (cyan $\circ$). Furthermore, we show numerical data for the
        CQTE with dimensions of a four-site Bose-Hubbard model with $N=15$
        particles (black $+$). In all cases COE statistics are used.
        The transition parameter is defined in
		Eq.~\eqref{transition_parameter}. The dashed vertical lines indicate
		the parameters (a-c) discussed in Sec.~\ref{sec:pheno_discussion}.
}
	\label{fig:entanglement_transition_w_resc}
\end{figure}

In Sec.~\ref{sec:transition_parameter}, we identify a rescaling of the
interaction strength $\epsilon$ into a dimension dependent transition
parameter $\Lambda(\epsilon, \{ k_{q_\uA}^\uA, k_{q_\uB}^\uB \})$. Considering
the interaction in scales of $\Lambda$, we obtain in
Fig.~\ref{fig:entanglement_transition_w_resc} that the number entropy and
total entropy behave universal for different choices of the
system dimensions. Therefore we show in
Fig.~\ref{fig:entanglement_transition_w_resc} the entropies for three different
combinations of dimensions, namely for $q=10, k=10$ and $q=50, k=5$ in the
equidimensional model and for a system with $N=15$ particles in the dimensions
of a four-site Bose-Hubbard model. For the number entropy and the total entropy,
Fig.~\ref{fig:entanglement_transition_w_resc}(i-ii), the curves fall after
rescaling with the RMT prediction \eqref{eq:rmt_S_2} and \eqref{eq:rmt__S_numb}
on top of each other, despite their large structural and dimensional
differences. This underlines that the physics within this model is universal
irrespective of the dimensions (as long as $q$ and $k$ are sufficiently
large). It motivates that an analytic description of the random matrix model
\eqref{eq:cqte_ensemble} can provide predictions for a large class of
deterministic models. Note that the configurational entropy in
Fig.~\ref{fig:entanglement_transition_w_resc}(iii) shows deviations. These are
due to the different scale of the visualization: From the RMT predictions
\eqref{eq:rmt_S_2} and \eqref{eq:rmt__S_numb}, we observe that
$S_\text{conf}^\infty \sim 1/q$ hence, $S_\text{conf}$ is typically
suppressed in the limit of many blocks in contrast to $S_2, S_\text{numb}
\sim 1$. By the rescaling of $S_\text{conf}$ fluctuations of order $1/q$
are thereby drastically enhanced and thus not comparable to the deviations
shown in Fig.~\ref{fig:entanglement_transition_w_resc}(i) and (ii).

\begin{figure}
	\includegraphics{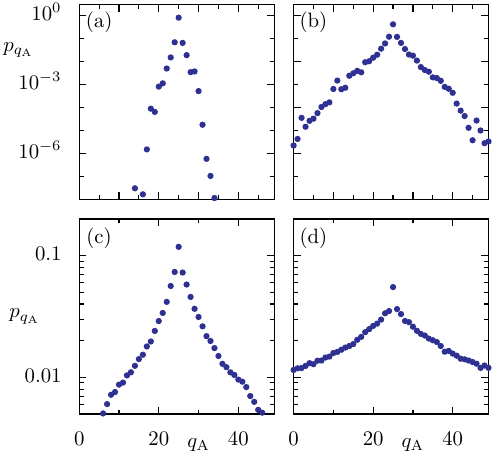}
	\caption{The block amplitudes for typical eigenstates of the CQTE with
     $q=50$ and $k=5$ and COE statistics averaged over all eigenstates which
	 have the maximal
     block amplitude in the block $q=25$ and in addition averaged over 5
     realizations of the CQTE. The choice of $\Lambda_q$ coincides with the
     parameters given in Fig.~\ref{fig:time_evolution_operator} and
     Fig.~\ref{fig:density_matrix_transition}.
     }
     \label{fig:localization_plot}
\end{figure}

Furthermore, we recall that the time-evolution operator possess a banded
matrix form. It is known that banded random matrices typically have
exponentially localized eigenstates, intensively investigated for Anderson
localization, see e.g.~\cite{Efe1983,FyoMir1994,AltZir1996}. Based on this we
expect the
eigenstates to exhibit a localization along the block diagonal, which can also
be seen in Fig.~\ref{fig:density_matrix_transition}. Furthermore, we depict in
Fig.~\ref{fig:localization_plot}(a-d) the block amplitudes $p_{q_\uA}$ for
various values of the coupling strength. These coupling strengths
are indicated in Fig.~\ref{fig:entanglement_transition_w_resc}
along the transition, corresponding to (a) weakly coupled
$\Lambda = 0.1$, (b) intermediate coupled $\Lambda=0.4$, (c) strongly
coupled parameters $\Lambda=1.2$ and (d) very strong coupling, $\Lambda=4$.
Each of these four parameters is discussed in the following by means of a
typical realization of a reduced density matrix and the block amplitudes
shown in Fig.~\ref{fig:density_matrix_transition}(a-d) and
Fig.~\ref{fig:localization_plot}(a-d), respectively.

(a) For small couplings the reduced density matrix is almost diagonal. As by
construction of the ensemble \eqref{eq:cqte_ensemble}, the reduced density
matrix in Fig.~\ref{fig:density_matrix_transition}(a) is given in the
eigenbasis of the uncoupled system. Hence we can conclude that the eigenbasis
of the uncoupled system yields a good approximation to the Schmidt states and
the Schmidt values are approximately given by the entries on the diagonal.
Note that the two largest diagonal elements are in neighboring blocks, which
we argue below is generic. The described observations become more pronounced
with decreasingly smaller couplings. Moreover, the states display an
exponential localization of their block amplitudes, see
Fig.~\ref{fig:localization_plot}(a).

(b) For the intermediate regime the reduced density matrix remains
approximately diagonal, hence the eigenstates of the uncoupled system still
approximate the Schmidt states. However, there are increasingly more non-zero
Schmidt values on the diagonal, contributing to a increasingly larger entropy.
The state is still localized but displays a broader width than in (a), and
the localization is no longer perfectly exponential. Here effects of the
boundary due to finite $q$ appear. Moreover, Schmidt values distribute over
different blocks, but also occur in the same block, generating configurational
entropy and hence entanglement between the subsystems.

(c) For strong coupling the reduced density matrix is no longer diagonal,
hence results about the Schmidt values cannot be deduced from the graphical
representation without diagonalization. The localization along the block
amplitudes persists but is not exponential. Instead there is a slow
convergence toward the random matrix distribution of the block values, which
is for the chosen structure the uniform distribution $p_\uA = 1/q$.

(d) For very strong coupling the eigenstates still show a deviation from the
expected uniform distribution in the block amplitudes $q_\uA$, see
Fig.~\ref{fig:localization_plot}(d). This manifests a robustness of
the localization mechanism and furthermore shows that the number entropy as a
measure of this localization tends rather slowly to the random matrix
prediction, suggesting a power law behavior. This is consistent with
expectations for banded random matrices \cite{FyoMir1994, AltZir1996}.

Most of above observations can be explained by the Fock-space local action
of the interaction: We consider a fixed eigenstate $\ket{ij; q_\uA}$ of the
initially uncoupled system. This eigenstate is not entangled, hence has a
single eigenvalue located in some subspace $\mathcal{H}^{(q_\uA)}_\uA$.
For small $\epsilon$ the coupling $\ue^{\ui V}$ is approximated by $1 + \ui
V$, which by definition \eqref{eq:coupling_potential} establishes correlations
between neighboring blocks of $q_\uA$ only, but not between subsystems $\uA$
and $\uB$ within the block of $q_\uA$. This absence of interaction and hence
entanglement within subsystems $\uA$ and $\uB$ in a fixed block motivates the
coincidence of number entropy and total entropy as $S_\text{conf} = 0$.
By proceeding to the second order in $\epsilon$, correlations of the form
$\sim V^2$ participate. These allow for ``back exchange processes''
$q_\uA \to q_\uA \pm 1 \to q_\uA$ and thereby generate entanglement in the
block-reduced density matrices $\tilde{\rho}_{q_\uA}$. Moreover, the processes
$q_\uA \to q_\uA \pm 2 \to q_\uA \pm 2$ establish non-zero coefficients
$p_{q_\uA \pm 2}$ contributing further to the number entropy.
The next higher order gives successively more non-vanishing contributions to
both the number entropy and the configurational entropy.

\subsection{Analytical derivation}

Having set up the required measures for quantifying the amount of entanglement
in a state, we start to quantitatively characterize the entanglement of
eigenstates generated by a varying coupling strength $\epsilon$. We
concentrate on the number entropy and the total entropy.
In order to describe this transition analytically we employ a perturbative
treatment in combination with an average over the statistical properties of
the random matrix system. This approach has proven to be a powerful analytical
tool in the setting without conservation laws \cite{SriTomLakKetBae2016,
LakSriKetBaeTom2016, TomLakSriBae2018, HerKieFriBae2020, PulLakSriBaeTom2020,
PulLakSriKieBaeTom2023}. We present here the required adaptions towards the
setting involving a conserved quantity.

\subsubsection{Perturbation theory} By implementing the notion of conserved
quantities explicitly into the framework of perturbation theory for unitary
systems \cite{TomLakSriBae2018}, we obtain for an eigenstate $\ket{\Psi_{jk};
q_\uA}$ of $\calU(\epsilon)$ in leading order of the interaction $V$
\begin{align}
	\ket{\Psi_{jk}; q_\uA} &\approx \tilde{n} \ket{jk; q_\uA}
		+ \ket{\Psi_{jk}^{(1)}; q_\uA} \label{eq:first_order_pert_series} \\
	\ket{\Psi_{jk}^{(1)}; q_\uA} &=
		\sum_{j'k';q_\uA' \neq jk;q_\uA}
			\frac{ \matrixel{j'k',q_\uA'}{V}{jk; q_\uA} }
			{\theta_{jk,q_\uA} - \theta_{j'k', q_\uA'}} \ket{j'k'; q_\uA'},
\end{align}
where $\ket{ij; q_\uA}$ is the eigenstate of the unperturbed system and $\{
\theta_{jk, q_\uA} \}$ the unperturbed eigenphases and
\begin{align}
 \tilde{n} &= 1 - \frac{1}{2}\sum_{j'k';q_\uA' \neq jk;q_\uA}
 \frac{|\matrixel{j'k',q_\uA'}{V}{jk; q_\uA}|^2}
 {(\theta_{jk,q_\uA} - \theta_{j'k', q_\uA'})^2},
\end{align}
is the normalization expanded to leading order. Note that the expansion is
effectively in orders  $V \sim \epsilon$. By definition
\eqref{eq:coupling_potential}, the coupling potential $V$ acts only on
neighboring pairs of subblocks and hence the leading order perturbation
simplifies as
\begin{align}
	\ket{\Psi_{jk}^{(1)}; q_\uA} &=
		\sum_{ \substack{j'k' \neq jk \\ s \in \{ +1, -1\}} }
			\frac{ h_{j'k', jk}^{(q_\uA + \iota(s))} }
			{\theta_{jk, q_\uA} - \theta_{j'k', q_\uA + s}}
			\ket{j'k'; q_\uA + s},
\end{align}
where $\iota(s) = (s+1)/2$. Again, the interaction strength $\epsilon$ enters
only implicitly through the variance of the random variables $h_{j'k',
jk}^{(q_\uA)}$.

\subsubsection{Statistical properties} In contrast to typical use-cases of
perturbation theory, the unperturbed eigenproblem is not analytically known.
Instead, the treatment relies on exploiting the statistical properties of the
uncoupled eigenproblem. By construction of the model, the subblock dynamics
are given as the tensor product of two independent random matrices. It is
known from random matrix theory, that in the limit of large matrix sizes the
eigenphases follow Poissonian statistics, i.e.,~are uncorrelated
\cite{TkoSmaKusZeiZyc2012}. In more detail, by rescaling the level differences
$s_{jk,j'k'} = (\theta_{jk} - \theta_{j'k'})/D$ with mean level spacing
\begin{align}
		D &= \frac{2 \pi}{M},
\end{align}
where $M$ is the matrix dimension, we have that the two point correlation
function
\begin{align}
	R_2(s) = \sum_{j'k'} \delta(s - s_{jk,j'k'}),
\end{align}
tend for $M \to \infty$ to the Poissonian statistics
\begin{align}
	R_2(s) &= 1. \label{eq:poission_stats}
\end{align}
We emphasize that these results hold only in the limit of large matrices and
therefore it is necessary to implement a well-behaved large matrix limit in
which the block dimensions become asymptotically large as well. Hence, we
introduce a parameter $M$ and impose
\begin{align}
		\tilde{k}^\uA_{q_\uA} &= k^\uA_{q_\uA} / \sqrt{M} = \text{const.} \\
		\tilde{k}^\uB_{q_\uB} &= k^\uB_{q_\uB} / \sqrt{M} = \text{const.},
\end{align}
i.e.,~all sectors increase simultaneously with $M\to \infty$, but their relative sizes
remain constant.

\subsubsection{Transition parameter} \label{sec:transition_parameter}
The large matrix limit has an important consequence on the formulation of
perturbation theory. As the mean level
spacing decreases for increasingly large matrices (due to the increased number
of eigenphases on the finite unit circle) the phase differences tend to
become resonant, i.e.,~$D \to 0$. More precisely and tailored for our
scenario: As the coupling establishes an interaction between neighboring
subblocks, only phase differences with levels from neighboring subblocks
arise in the sum. Their number is $k_{q_\uA + 1}$ and $k_{q_\uA - 1}$,
respectively, giving a mean level spacing
\begin{align}
	D_{q_\uA \pm 1} &= \frac{2 \pi}
			{M \tilde{k}_{q_\uA \pm 1}}.
\end{align}
To avoid the implicitly dimension-dependent character of the level spacing, we
introduce a normalization such that
\begin{align}
	\theta_{jk} - \theta_{j'k'} = D_{q_\uA \pm 1} s_{jk, j'k'}
\end{align}
where the fluctuations $s_{jk, j'k'}$ have unit mean-value, i.e.,~remain
finite in the limit $M \to \infty$.
As we average over the absolute square of $V$, the average scaling is given by
the variance \eqref{eq:coupling_variance} such that
\begin{align}
	|h^{q_\uA}_{j'k', jk}|^2 = \beta \sigma_{q_\uA}^2 \omega_{jk,j'k'},
\end{align}
where fluctuations $\omega_{jk,j'k'}$ are again effectively random variables
with unit mean value distributed according to
\begin{align}
   \varrho(w) &= \begin{cases}
                      \ue^{-w} & \text{CUE} \\
                      \frac{\ue^{-w/2}}{\sqrt{2\pi w}} & \text{COE}
                     \end{cases}.
\end{align}
Thus they follow the exponential or Thomas-Porter distribution for the unitary
or orthogonal symmetry class, respectively. These distributions arise as the
absolute square of a complex Gaussian or real Gaussian distribution. Combining
the mean values of level spacing $D_{q_\uA}$ and matrix element $\beta
\sigma_{q_\uA}^2$, we can define a single universal parameter
$\Lambda_{q_\uA}$ explicitly given by
\begin{align}
	\sqrt{\Lambda_{q_\uA}} &= \sqrt{\beta}
		\biggr( \frac{ \sigma_{q_\uA+1}}{D_{q_\uA+1}} +  \frac{
\sigma_{q_\uA}}{D_{q_\uA-1}} \biggr)
				\label{eq:gen_block_transition_parameter} \\
			&= \frac{\epsilon}{\pi}
			\left(
			\frac{\tilde{\sigma}_{q_\uA+1}}{2}
			  \sqrt{\frac{\tilde{k}_{q_\uA + 1}}{\tilde{k}_{q_\uA}}}
		  + \frac{\tilde{\sigma}_{q_\uA}}{2}
				\sqrt{\frac{\tilde{k}_{q_\uA -1}}
					{\tilde{k}_{q_\uA}}} \right).
		\label{eq:cqte_transition_parameter}
\end{align}
We note that the variance \eqref{eq:coupling_variance} is chosen in advance to
generate an asymptotically stable combination which is independent of $M$.
Importantly, this parameter imposes the universality of the system: the
phenomenological effects are no longer depending on the $\epsilon$ or the
dimensions $\{ \tilde{k}_{q_\uA} \}$, but on its combination. Hence, a
different choice of $\tilde{\sigma}_{q_\uA}$ leaves the phenomenological small
coupling behavior unchanged as long as $\epsilon$ is adjusted to keep $\Lambda$
constant. This universality holds beyond the perturbative regime as is
numerically illustrated in Fig.~\ref{fig:entanglement_transition_w_resc}.

\subsubsection{Reduced density matrix and Schmidt values}

We can now turn towards the characterization of the Schmidt values of the
reduced density matrix. We have numerically observed in
Fig.~\ref{fig:density_matrix_transition}(a-b) that for small couplings, the
reduced density matrix remains essentially diagonal. Hence the Schmidt values
can be approximately identified as the leading diagonal entries of the
perturbation expansion. The normalization in front of the unperturbed
eigenstate in \eqref{eq:first_order_pert_series} is close to unity and can be
identified as the largest Schmidt value, i.e.,
\begin{align}\label{Schmidt value}
	\lambda_1^{jk}  = 1
	& -\frac{\beta\sigma_{q_\uA+1}^2}{D_{q_\uA+1}^2}
		\sum_{\substack{j'k' \\ q_\uA' = q_\uA + 1}}
			\frac{w_{jk,j'k'}}{s_{jk,j'k'}^2} \nonumber \\
	&- \frac{\beta\sigma_{q_\uA}^2}{D_{q_\uA-1}^2}
		\sum_{\substack{j'k' \\ q_\uA' = q_\uA - 1}}
			\frac{w_{jk,j'k'}}{s_{jk,j'k'}^2}.
\end{align}
This result characterizes the first Schmidt value of a fixed eigenstate of a
fixed realization of the random matrix ensemble. In order to implement the
ensemble average, we first replace the sum by an integral such that
\begin{equation}
	\sum_{\substack{j'k' \\ q_\uA' = q_\uA \pm 1}}
	\frac{w_{jk,j'k'}}{s_{jk,j'k'}^2} =
		\int_{0}^{\infty} \ud w \int_{-\infty}^{\infty} \ud s
			~\frac{w}{s^2} R^{jk,(\pm)}_{q_A}(s, w),
\end{equation}
where we define
\begin{equation}
	R_{q_\uA}^{jk,(\pm)}(s, w) =
		\sum_{\substack{j'k' \\ q_\uA' = q_\uA \pm 1}}
			\delta(w - w_{jk,j'k'})\delta(s-s_{jk,j'k'}).
\end{equation}
The average over realizations of the CQTE acts now on the distribution
$R_{q_\uA}^{jk,(\pm)}(s, w)$ only. We assume, typical for random matrix theory,
that the eigenphase statistics and eigenstate statistics are independent and
uncorrelated from each other such that
\begin{equation}
	\overline{R_{q_\uA}^{jk,(\pm)}(s, w)} = R_2(s) \varrho(w),
\end{equation}
where $R_2(s)$ follows Poissonian statistics \eqref{eq:poission_stats}.
Collecting the results, we obtain the integral
\begin{equation}
	\lambda_1^{q_A} = 1 - \biggr(
		\frac{\beta \sigma_{q_\uA+1}^2}{D_{q_\uA+1}^2}
	  + \frac{\beta \sigma_{q_\uA}^2}{D_{q_\uA-1}^2} \biggr)
		\int_{0}^{\infty} \ud w \int_{-\infty}^{\infty} \ud s
			~ \frac{w}{s^2} \varrho(w),
	\label{eq:lamb_1_unreg_int}
\end{equation}
for the largest Schmidt value. This integral is divergent as $\frac{w}{s^2}$
has a non-integrable pole at $s = 0$. This is due to frequently arising
resonances of a Poissonian spectrum, i.e.,~the lack of repelling in
uncorrelated levels distributions.

\subsubsection{Regularization of resonances}
We assume in the presence of interactions, that these level crossings
immediately start to repel in the form of avoided level crossings. This
behavior can be implemented in form of a regularization
\cite{LakSriKetBaeTom2016, TomLakSriBae2018}, giving rise to the replacement
\begin{align}
&\frac{|h_{j'k',jk}^{q_\uA}|^2}{(\theta_{j'k',q_\uA'}-\theta_{jk,q_\uA})^2}
	\rightarrow \nonumber \\
	&\quad \frac{1}{2}\biggr[ 1 - \biggr(1+4\frac{|h_{j'k',jk}^{q_\uA}|^2}
	{(\theta_{j'k',q_\uA'}-\theta_{jk,q_\uA})^2}
		\biggr)^{-1/2}\biggr],
\end{align}
which is obtained by treating levels with $\theta_{j'k',q_\uA'}
- \theta_{jk,q_\uA} \approx 0$ as degenerate and using perturbation theory for
degenerate spectra. Note that this regularization applies separately to the
two terms in \eqref{eq:lamb_1_unreg_int}.

\begin{figure}\label{fig:Universal_schmidt}
	\includegraphics{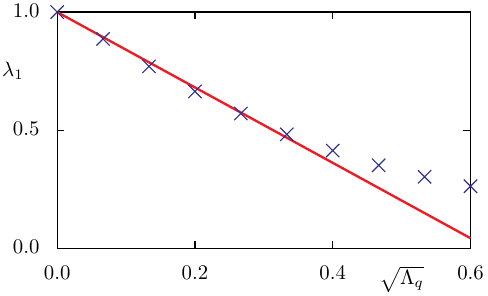}
	\caption{Average of the largest Schmidt eigenvalue as a function of
	$\sqrt{\Lambda_{q}}$. The numerical data (blue $\cross$) corresponds to
	the CQTE with the subsystems chosen from the COE. The red line is
	the perturbative result, Eq.~\eqref{transition_total_schmidt}.
        }
\end{figure}

To this end, the integral of the leading Schmidt value converges by means of
the regularization and gives rise to
\begin{equation}\label{Schmidt Trans. Block}
	\lambda_1^{q_\uA} = 1 - \sqrt{\Lambda_{q_\uA}}\begin{cases}
	     \sqrt{\pi} & \text{CUE} \\
		\frac{4}{\sqrt{2\pi}} & \text{COE}.
	\end{cases}
\end{equation}
This approximation of the Schmidt values applies to an average over
eigenstates localized within a fixed block $q_\uA$.
In order to obtain the average transition of the largest Schmidt value for all
eigenvectors, we have to average over all values of $q_\uA$ weighted by the
multiplicity $k_{q_\uA}$ of states with a fixed $q_\uA$. Thus, the first
Schmidt value averaged over all eigenstates is given by
\begin{align}\label{transition_total_schmidt}
	\overline{\lambda_1}
		= \frac{1}{K} \sum_{q_\uA} k_{q_\uA} \lambda_1^{q_\uA}
		= 1 - \sqrt{\Lambda_q}
			\begin{cases}
				\sqrt{\pi} & \text{CUE} \\
				\frac{4}{\sqrt{2\pi}} & \text{COE}.
			\end{cases}
\end{align}
Here the total transition parameter $\Lambda_q$ is defined by its square
root as
\begin{equation}\label{transition_parameter}
	\sqrt{\Lambda_q} =
		\frac{1}{K} \sum_{q_\uA} k_{q_\uA} \sqrt{\Lambda_{q_\uA}}.
\end{equation}
In Fig.~\ref{fig:Universal_schmidt} we observe that the transition of
$\lambda_1$ is in good agreement with the analytical result given in
Eq.~\eqref{transition_total_schmidt} for sufficiently small $\sqrt{\Lambda_q}$.

\subsubsection{Moments of the Schmidt values}\label{sec:moments_of_schmidt_values}

The remainder of the perturbative approach, i.e., the derivation of the linear
entropy from the perturbative expression follows without significant
modification the derivation of Refs.~\cite{HerKieFriBae2020,
TomLakSriBae2018}. We therefore state the main results directly without
repeating the detailed derivation.

We obtain for the sum over the square of all Schmidt values $\lambda_j$ with
$j > 1$ as
\begin{align}
	\overline{\sum_{j > 1} \lambda_j^2} &=
	\sqrt{\Lambda_{q_\uA}}\begin{cases}
		(1-\pi/4)\sqrt{\pi} & \text{CUE} \\
		\frac{4-\pi}{\sqrt{2\pi}} & \text{COE},
	\end{cases}
	\label{eq:2nd_moment_lambda_1}
\end{align}
and for the square of the largest Schmidt value
\begin{align}
	\overline{\lambda_1^2}
	&=  1 - \sqrt{\Lambda_{q_\uA}}\begin{cases}
		(1+\pi/4)\sqrt{\pi} & \text{CUE} \\
		 \frac{4+\pi}{\sqrt{2\pi}} & \text{COE}
	\end{cases}.
	\label{eq:2nd_moment_lambda_j}
\end{align}
Again, these results correspond to eigenstates with a fixed value of $q_\uA$.
In Ref.~\cite{TomLakSriBae2018} it is shown that in leading order the moments
$\overline{\sum_{j > 1} \lambda_j^\alpha}$ are equal to the moments of the
second largest Schmidt value $\overline{\lambda_2^\alpha}$. Therefore,
all $\lambda_j$ for $j \geq 3$ only contribute in higher order of $\Lambda_q$,
which will be crucial when later giving a description of the complete
transition.

For the entanglement entropy restricted to states with value $q_\uA$, combining
Eq.~\eqref{eq:2nd_moment_lambda_1} and Eq.~\eqref{eq:2nd_moment_lambda_j} we
obtain an average value of
\begin{equation}
	\overline{S_{2,q_\uA}} = 1 - \sum_{j} \overline{\lambda_j^2} =
\sqrt{\Lambda_{q_A}}\begin{cases}
	\pi^\frac{3}{2}/2 & \text{CUE} \\
	\sqrt{2\pi} & \text{COE},
\end{cases}
\end{equation}
for small $\Lambda_{q_\uA}$. The average linear entanglement entropy for all
values of $q_\uA$ is ultimately given by
\begin{equation}\label{perturbative entropy}
	\overline{S_{2}} = \frac{1}{K}
							\sum_{q_\uA}k_{q_\uA}\overline{S_{2,q_\uA}}
					 = \sqrt{\Lambda_{q}}\begin{cases}
					 	\pi^\frac{3}{2}/2 & \text{CUE} \\
					 	\sqrt{2\pi} & \text{COE}.
					 \end{cases}
\end{equation}
We emphasize that by our previous argumentation in
Sec.~\ref{sec:pheno_discussion} the total entropy coincides with the number
entropy. In the following we discuss a recursive procedure to obtain
non-perturbative predictions of the total entropy $S_2$.

\subsubsection{Recursive extrapolation}
\label{sec:recursive_extrapolation}

In the previous section 
we have obtained a perturbative description of the entanglement transition suited for
small values of the transition parameter. In contrast, the numerical
investigations, see Fig.~\ref{fig:entanglement_transition_w_resc},
demonstrate that the universality with $\Lambda_q$ remains far beyond the
perturbative regime. This suggests that the
perturbation theory can be pushed to some extent into a convergent
perturbation series giving the analytic expansion of a non-perturbative
result. At the same time, the direct implementation of higher order terms of
the perturbation theory is limited as they involved more complicated
information about avoided crossings of increasingly more levels. Instead, we
give a heuristic description based on ideas of the recursive embedded
perturbation theory presented in \cite{TomLakSriBae2018}. The central aspect
for this approach is the emergent cascade of Schmidt values for increasing
interaction. However, in order to describe the transition in systems with a
conserved quantity this recursive description has to be modified. Furthermore,
we note that this Ansatz is limited to the linear entropy $S_2$ and does not
apply for the number entropy.

\begin{figure}
	\includegraphics{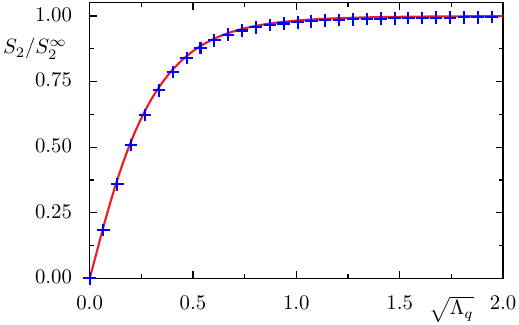}
	\caption{Linear entanglement entropy $S_2$ as a function of
	$\sqrt{\Lambda_{q}}$. The numerical data corresponds to the CQTE with COE
	statistics, where $k=k_{q_\uA}^\uA=k_{q_\uB}^\uB=10$ and $q = 36$ (blue
	$+$).
	The red line corresponds to the analytic result,
	Eq.~\eqref{eq:recursive_result_entanglement_entropy_w._asymptotic_val}.
	All curves are divided by their asymptotic value $S_2^\infty$.}
	\label{fig:rec_transition_cqte_coe}
\end{figure}

\begin{figure}
\includegraphics{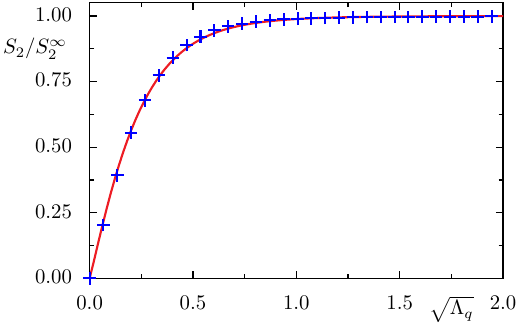}
	\caption{Same as Fig.~\ref{fig:rec_transition_cqte_coe} but for
	the CUE random matrix transition ensemble.}
	\label{fig:rec_transition_cqte_cue}
\end{figure}

In the following discussion we make the coupling strength $\epsilon$ more
explicit by considering a coupling operator
\begin{equation}
	\mathcal{U}_{\uA\uB}^{(q)}(\epsilon) = \ue^{\ui \epsilon V},
\end{equation}
where $V$ is chosen as in Sec.~\ref{subsec:dynamcis} but with a
standard deviation which is independent of the coupling strength.
First, we consider eigenstates $\ket{a(\epsilon)}$ of $\mathcal{U}(\epsilon)$.
The eigenstates for a slightly increased coupling strength $\epsilon + \Delta
\epsilon$ are then given by
\begin{align}\label{recursive_transition}
	\ket{a(\epsilon + \Delta \epsilon)} &=
	\left( 1 - \frac{\Delta \epsilon^2}{2} \sum_{b \neq a}
	\frac{|\matrixel{a(\epsilon)}{V}{b(\epsilon)}|^2}
	{(\theta_{a(\epsilon)} - \theta_{b(\epsilon)})^2} \right)
	\ket{a(\epsilon)} \nonumber \\
	&\qquad + \Delta \epsilon \sum_{b \neq a}
	\frac{\matrixel{b(\epsilon)}{V}{a(\epsilon)}}
	{\theta_{a(\epsilon)}-\theta_{b(\epsilon)}}
	\ket{b(\epsilon)},
\end{align}
where the sum runs over all eigenstates $\ket{b(\epsilon)}$ different
from $\ket{a(\epsilon)}$. This allows to determine the difference
of the average purity $\overline{\mu}$ of the eigenstates at
coupling strengths $\epsilon$ and $\epsilon + \Delta \epsilon$ as
\begin{equation}
	\overline{\mu(\epsilon + \Delta \epsilon)} - \overline{\mu(\epsilon)}
	= -\left( 1  - v^2(\epsilon,\Delta \epsilon)
				- v'^2(\epsilon, \Delta \epsilon)
	  \right)
	  \overline{\mu(\epsilon)},
       \label{eq:discrete-diff-eq-for-mu}
\end{equation}
where
\begin{align}
	v^2(\epsilon, \Delta \epsilon) &=
	    \overline{\left(1 - \Delta \epsilon^2 \sum_{b \neq a}
			\frac{|\matrixel{a(\epsilon)}{V}{b(\epsilon)}|^2}
				 {(\theta_{a(\epsilon)} - \theta_{b(\epsilon)})^2}
				  \right)^2} \label{eq:v_recursive}\\
	v'^2(\epsilon,\Delta \epsilon) &=
		\overline{\sum_{b \neq a} \left( \Delta \epsilon^2
			\frac{|\matrixel{b(\epsilon)}{V}{a(\epsilon)}|^2}
				 {(\theta_{a(\epsilon)}-\theta_{b(\epsilon)})^2}
			\right)^2}. \label{eq:v_prime_recursive}
\end{align}
A detailed derivation is given in App.~\ref{app:recursive_emmbedding}.
The challenge is to compute $v^2$ and $v'^2$ for arbitrary $\epsilon$. This
would allow to turn \eqref{eq:discrete-diff-eq-for-mu} into a differential
equation for $\overline{\mu}$, whose solution approximately describes the
entropy transition in a non-perturbative regime, see
Ref.~\cite{TomLakSriBae2018}. However this computation can become arbitrarily
complicated, as one needs to know the matrix element distribution and level
spacing distribution for all coupling strengths $\epsilon$. Nevertheless, the
observations presented earlier in this section allow for deducing important
implications for the behavior of $v^2$ and $v'^2$.
We observe in Fig.~\ref{fig:localization_plot} that the eigenstates are
localized for $\epsilon = 0$ and delocalize towards the RMT distribution for
increasing $\epsilon$. The coupling matrix $V$ only couples neighboring
sectors of the conserved quantity $Q_\uA$. Hence, the more delocalized
the eigenstates $\ket{a(\epsilon)}$ become, the more matrix elements
$|\matrixel{b(\epsilon)}{V}{a(\epsilon)}|^2$ turn non-zero. We therefore
conclude that the sum in Eq.~\eqref{eq:v_recursive}
and~\eqref{eq:v_prime_recursive} increases with $\epsilon$. For small $\Delta
\epsilon$, this causes the prefactor in Eq.~\eqref{eq:discrete-diff-eq-for-mu},
$1-v^2-v'^2$, to increase with $\epsilon$, resulting in a
faster-than-exponential decay of the purity. The transition from localization
to delocalization of the eigenstates is a direct consequence of the breaking
of the conserved quantity. Therefore this is absent in systems without
conserved quantities, see Ref.~\cite{TomLakSriBae2018}. In this case $v^2$ and
$v'^2$ can be approximated as constant with respect to $\epsilon$, yielding an
exponential decay of the purity. In App.~\ref{app:recursive_emmbedding} we
present an approach which incorporates the observed delocalization for yet
small values of $\epsilon$. For eigenstates in the COE case which correspond
to a fixed sector $q_\uA$ for $\epsilon = 0$, the purity is then given by
\begin{align}
	\overline{\mu_{q_\uA}}(\Lambda_{q_\uA}) &\approx
			\ue^{-s(\Lambda_{q_\uA})},
\end{align}
where
\begin{align}
	s(\Lambda_{q_\uA}) &=
		  \sqrt{2\pi} \sqrt{\Lambda_{q_\uA}}
		+ (2\pi)^{1/4}\frac{3\sqrt{2}+2}{3}\Lambda_{q_\uA}^{3/4}
		- \frac{3}{2} \Lambda_{q_\uA},
\end{align}
and $\Lambda_{q_\uA}$ is the transition parameter defined in
Eq.~\eqref{eq:cqte_transition_parameter}. The results for the CUE case are given
in App.~\ref{app:recursive_emmbedding}. By averaging over all values of $q_\uA$
we obtain the linear entanglement entropy
\begin{equation}
	\overline{S_2} =
		1 - \frac{1}{K} \sum_{q_\uA} k_{q_\uA}
			\overline{\mu_{q_\uA}}(\Lambda_{q_\uA}).
	\label{eq:recursive_result_entanglement_entropy}
\end{equation}
A simple interpolation between
Eq.~\eqref{eq:recursive_result_entanglement_entropy} and the asymptotic value,
Eq.~\eqref{eq:rmt_S_2}, gives
\begin{equation}
	\overline{S_2} = S_2^\infty
		\left( 1 - \frac{1}{K} \sum_{q_\uA} k_{q_\uA}
			\overline{\nu_{q_\uA}}(\Lambda_{q_\uA})
			\right),
	\label{eq:recursive_result_entanglement_entropy_w._asymptotic_val}
\end{equation}
where $\overline{\nu_{q_\uA}} = (\overline{\mu_{q_\uA}})^{1/S_2^\infty}$.
Note that the sum in Eq.~\eqref{eq:recursive_result_entanglement_entropy_w._asymptotic_val}
can still depend on higher order contributions in
$\sqrt{\Lambda_{q_\uA}}$, which cannot be collected into a single overall
transition parameter $\sqrt{\Lambda_q}$. As a consequence there is still a
dependence on $q$, which, however is very small for large systems and can be
practically neglected. Furthermore, we observe that $s(\Lambda_{q_\uA})$ is eventually decreasing
for large $\Lambda_{q_\uA}$, which leads to $S_2$ decreasing away from its
asymptotic value. This effect results from the fact that the delocalization of
$\ket{a(\Lambda_{q_\uA})}$ is accounted for only up to first order and should
be absent in more sophisticated approximations. However, this effect occurs
only for values of $\Lambda_{q}$ that are substantially larger than the
physically interesting range in which the transition takes place.
A comparison with the numeric data of the CQTE is shown in
Fig.~\ref{fig:rec_transition_cqte_coe} for the COE and in
Fig.~\ref{fig:rec_transition_cqte_cue} for the CUE. Although the respective
calculations in App.~\ref{app:recursive_emmbedding} are carried out for small
$\Lambda_{q_\uA}$ ($\Lambda_q$), they provide a good agreement over
the entire transition.

\section{The kicked Bose-Hubbard model}
\label{sec:Bose_Hubbard_model}
The perturbative approach and the recursive description obtained in the
previous sections are well-suited for the random matrix transition ensemble.
In the remainder we investigate how these results apply to physical systems.

An important system with a conserved quantity is the kicked Bose-Hubbard model
\cite{StrGraKor2008, FavFazRus2020}. It describes spinless bosons on a lattice
and arises in the study of optical lattices and Bose-Einstein condensates. The
general kicked Bose-Hubbard Hamiltonian is given by
\begin{align}
	H(t) &= H_\text{lin}
			+ H_{\text{kick}}\sum_{m=-\infty}^\infty \delta(t-m),
			\label{eq:h_tot} \\
	H_\text{lin} &= -\sum_{\ell=1}^L J_{\ell} (a_\ell^{\dagger}a_{\ell+1})
			+ \text{h.c.}, \label{eq:h_lin}  \\
	H_\text{kick} &= \frac{U}{2}\sum_{\ell=1}^L n_\ell(n_\ell-1)
			+ \sum_{\ell=1}^L\mu_\ell n_\ell, \label{eq:h_kick}
\end{align}
where $a_i, a_j^\dagger$ are bosonic creation and annihilation operators,
i.e.,~$[a_i, a_j^\dagger] = \delta_{ij}$, and $n_\ell = a_\ell^\dagger a_\ell$
is the particle number operator on site $\ell$. The system consists of $N$
particles and $L$ sites, where we assume periodic boundary conditions, i.e.,~$L
+ 1 \equiv 1$. The parameters are the local on-site interaction $U > 0$,
the coupling strength $J_{\ell} > 0$ between neighboring sites $\ell$ and
$\ell+1$ and the local on-site potentials $\mu_\ell$, which establish a
disorder allowing to break discrete symmetries. The unitary time evolution
operator is given by
\begin{align}
	\mathcal{U}
	&= \ue^{-\frac{\ui}{\hbar} H_\text{lin}}
			\ue^{-\frac{\ui}{\hbar} H_\text{kick}}.
		\label{time-evolution}
\end{align}
We assume $\hbar = 1$ in the remainder. The system posses a classical
(mean-field) limit for $N \rightarrow \infty$ while $U \rightarrow 0$ and
keeping $\kappa = UN$ fixed \cite{Kol2016, DubMue2016, RicUrbTom2022}. This
scaling ensures that the contributions from $H_\text{lin}$ and $H_\text{kick}$
stay comparable in orders of $N$.

In order to study the entanglement properties of the kicked Bose-Hubbard model
we need to clarify the bipartite structure. A natural bipartition is obtained
by dividing the system in real space. We define a subsystem $\uA$ to consist
of sites $1 \leq i \leq L_\uA$ and subsystem $\uB$ of sites $L_\uA+1 \leq i
\leq L$. We consider $L_\uA = L_\uB = 2$ as this is the simplest case showing
chaotic dynamics within each subsystem \cite{StrGraKor2008}. The interaction
between different sites is generated by the coupling terms $J_\ell
a_\ell^{\dagger} a_{\ell + 1} + \text{h.c.}$, see Eq.~\eqref{eq:h_lin}. In
particular, the interaction between the subsystems $\uA$ and $\uB$ is given by
the coupling terms for $\ell=2$ and $\ell=4$. All other terms in the
Hamiltonian \eqref{eq:h_tot} contribute to the dynamics within the subsystems.
We choose $J_1 = J_3 = J$ and $J_2 = J_4 = \epsilon$, such that the kinetic
part of the Hamiltonian $H_\text{kin}$ is given by
\begin{align}
	H_\text{lin} &= - J(a_1^{\dagger}a_2 + a_3^{\dagger}a_4)
					- \epsilon (a_2^{\dagger}a_3 + a_1^{\dagger}a_4)
					+ \text{h.c.} \\
				&=: H_{\text{lin} , 0} - \epsilon V.
\end{align}
The parameter $\epsilon$ governs the coupling between the subsystems, such
that the subsystems $\uA$ and $\uB$ are non-interacting for $\epsilon = 0$,
while we obtain a fully coupled Hamiltonian for $\epsilon \sim J$. To choose
suitable parameters we start with the uncoupled case and consider a pair of
sites with $N$ particles. For $J = \pi/4$, $\kappa=48$, and uniformly
distributed $\mu_\ell \in [-1/2, 1/2]$ the consecutive level spacing
distribution is found to be well described by the COE. Thus each subsystem
qualifies as quantum chaotic. Coupling these two subsystems leads to an
entanglement transition which should be well described by the CQTE.

To apply the theory we first bring the time evolution operator
\eqref{time-evolution} into the bipartite form
\begin{equation}
	\mathcal{U}(\epsilon)
		= \mathcal{U}_{\text{AB}}(\epsilon)\mathcal{U}_0
		= \ue^{\ui \epsilon V} \ue^{-\ui H_{\text{lin},0}}
			\ue^{-\ui H_\text{kick}},
\end{equation}
where $\mathcal{U}_{\text{AB}}(\epsilon) = \ue^{\ui \epsilon V}$ and
$\mathcal{U}_0 = \ue^{-\ui H_{\text{lin},0}}\ue^{-\ui H_{0} }$
and $[H_{\text{lin},0}, V] = 0$ has been used. The conserved quantity $Q$ is
given by the total particle number operator $\mathcal{N} = \mathcal{N}_\uA +
\mathcal{N}_\uB$. Consequently we have $[\mathcal{U}(\epsilon), \mathcal{N}] =
0$ and $[\mathcal{U}_0, \mathcal{N}_\uA] = [\mathcal{U}_0, \mathcal{N}_B] = 0$.
The interaction operator $V = a_2^{\dagger} a_3 + a_1^{\dagger}a_4 +
\text{h.c.}$ creates and annihilates only one particle in subsystem $\uA$ and
$\uB$, respectively. The non-interacting time evolution operator splits into a
sum over all subsystem particle numbers,
\begin{equation}
 	\mathcal{U}_0 = \sum_{N_\uA=0}^N
		\mathcal{U}^{\uA}_{N_\uA} \otimes \mathcal{U}^{\uB}_{N-N_\uA},
		\label{time evolution BH}
\end{equation}
where $\mathcal{U}^{\uA}_{N_\uA}$ is the projection of
\begin{equation}
	\ue^{-\ui J(-a^\dagger_1 a_2 - a^\dagger_2 a_1)}
	\ue^{-\ui \sum_{\ell= 1,2} \frac{U}{2} n_\ell(n_\ell+1)
		+ \mu_\ell n_\ell}
\end{equation}
to a fixed value $N_\uA$ and $\mathcal{U}^{\uB}_{N_\uB}$ is defined in a
similar way with respect to site $3$ and $4$. These operators,
$\mathcal{U}^{\uA}_{N_\uA}$ and $\mathcal{U}^{\uB}_{N_\uB}$, have a
time-reversal symmetry, so that they correspond to the CQTE
\eqref{eq:cqte_ensemble} based on COE statistics. In order to obtain the
transition parameters $\Lambda_{N_\uA}$,
Eq.~\eqref{eq:gen_block_transition_parameter}, we have to replace the
variance by the mean squared off-diagonal matrix elements
\begin{equation}
	\sigma^2_{N_\uA} = \epsilon^2\overline{|\matrixel{j'k';N_\uA'}{V}{jk;N_\uA}|^2},
\end{equation}
for random COE vectors $\ket{j}$, $\ket{j'}$, $\ket{k}$ and $\ket{k'}$ with
$N_\uA' = N_\uA -1$. This evaluates to
\begin{equation}
	\tilde{\sigma}_{N_\uA}^2 = \frac{k_{N_\uA}k_{N_\uA-1}}{2}.
\end{equation}
Hence, the subsystem transition parameter
$\Lambda_{N_\uA}$ is given by
\begin{equation}
	\Lambda_{N_\uA} = \frac{(k_{N_\uA-1}+k_{N_\uA+1})^2}{8\pi^2}\epsilon^2,
	\label{transition parameter N_A}
\end{equation}
where $k_{N_\uA} = (N_\uA+1)(N-N_\uA+1)$ is the dimension of the subspace with
fixed subsystem particle number $N_\uA$. Averaging over all subsystem particle
numbers $N_\uA$ using \eqref{transition_parameter} yields the total transition
parameter
\begin{equation}\label{transition parameter N}
	\Lambda_N = \frac{N^2 (N + 4)^2}{50\pi^2}\epsilon^2.
\end{equation}

\begin{figure}
\includegraphics{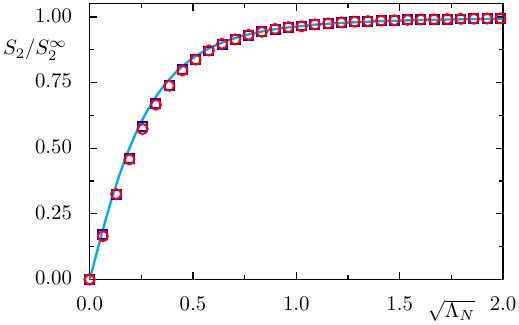}
	\caption{Average of the linear entanglement entropy $S_2$ for the kicked
	Bose-Hubbard model with $L=4$ and $N=20$ (blue
    $\square$) and $N = 24$ (red $\circ$) as a function of
    $\sqrt{\Lambda_{N}}$.
    The parameters are $\kappa = 48$ and $J = \pi/4$. The data is averaged over $10$
    realizations of $\mu_\ell$. The cyan line indicates the analytical result
    \eqref{eq:recursive_result_entanglement_entropy_w._asymptotic_val} for $N=24$. All curves are divided by their asymptotic value $S_2^\infty$.
}
\label{fig:rec_transition_bose_hubbard}
\end{figure}

Figure~\ref{fig:rec_transition_bose_hubbard} shows a comparison of the
linear entanglement entropy of the kicked Bose-Hubbard model and the analytical
result \eqref{eq:recursive_result_entanglement_entropy_w._asymptotic_val},
using the transition parameter \eqref{transition parameter N}. The
analytical result describes the numerical data well.

We conclude the discussion by considering the behavior for smaller values of
$\kappa$. In this case we find that $S_2(\sqrt{\Lambda_N})$ shows a small
dependence on the value of the conserved quantity $N$ (not shown). The origin
of these deviations from the analytical result \eqref{eq:recursive_result_entanglement_entropy_w._asymptotic_val} lies in the structure of $\mathcal{U}(\epsilon)$, see Eq.~\eqref{time
evolution BH}. It contains operators $\mathcal{U}^{\uA}_{N_\uA}$ for all $0
\leq N_\uA \leq N$. These operators describe the two-site dynamics with
$N_\uA$ particles. These are governed by the effective parameter $\tilde{\kappa}
= U N_\uA = \kappa N_\uA / N$. For small $N_\uA/N$ this leads to operators
$\mathcal{U}^{\uA}_{N_\uA}$ (and analogously for $\mathcal{U}^{\uB}_{N_\uB}$)
whose consecutive level spacing becomes closer to Poissonian statistics
indicative for integrable systems. Therefore the underlying assumption of the
description by the random matrix transition ensemble is not fulfilled for all
terms of the time evolution operator \eqref{time evolution BH}. It is worth
emphasizing that the fraction of these operators with small effective
$\tilde{\kappa}$ does not diminish when increasing $N$. Still the random
matrix transition ensemble provides a good description of the entanglement in
the kicked Bose-Hubbard model as the dominant contributions to the entanglement entropy
come from subspaces with sufficiently large effective $\tilde{\kappa}$.

\section{Summary and outlook}
\label{sec:summary}

In this paper we study the entanglement generation in a wide range of
bipartite systems showing a break up of conservation laws. We assume that in
the non-interacting case, each subsystem behaves quantum chaotically and
posses an individual conserved quantity. Furthermore we suppose that
interactions between the subsystems break the individual conservation laws
while the corresponding global quantity remains conserved. We propose a
suitable random matrix ensemble, which allows to study
the universal transition from uncoupled to strongly interacting subsystems
independent of the specific underlying dynamics.
We observe that the conserved quantities impose a block structure on the
time-evolution operator. This structure is taken into account by symmetry
resolved entanglement, a general
framework to study entanglement in the presence of a conserved quantity. This
leads to the entropy splitting into two parts, the number entropy and the block
entropy. The number entropy can be understood as correlations that originate
from the conserved quantity, whereas the block entropy corresponds to standard
quantum mechanical correlations between subsystem.

Based on this framework, we find that the  transition of entanglement
is governed by a universal transition parameter $\Lambda$,
which only depends on the system size of the involved sector and the
interaction strength between the subsystems. For weakly coupled
subsystems, we observe a localization along the sectors of the broken
individual conserved quantity. By increasing interaction strength the
eigenstates delocalize towards the fully chaotic distribution predicted by
random matrix theory. In the weakly coupled regime, we observe that the
entanglement entropy is completely determined by the number entropy. This is
due to the structure of the interaction, which couples neighboring sectors of
conserved quantity only.  Hence, for weak coupling, the entanglement
originates from the correlations of the Fock-space structure and only
implicitly by the explicit subsystems.
Furthermore, we analytically derive the leading order expansions of the
largest Schmidt values, the number entropy, and the block entropy.
We give a description of the non-perturbative regime based on heuristic
arguments similar to Ref.~\cite{LakSriKetBaeTom2016, TomLakSriBae2018}. We
find good agreement with numerical data. We apply the result to the kicked
Bose-Hubbard systems, which is an explicit example of a system with the
considered structure of  conservation laws. We again find good agreement with
the analytic prediction results. This underlines the universality of the
entanglement transition, i.e.,~its independence of the specific system
dynamics.

A future direction is the investigation of entanglement generation via
time-evolution within this setting. Furthermore, it will be
interesting to consider adaptions to the case of conserved
quantities arising from non-abelian symmetries \cite{BiaDonKum2024} such as
spin conservation.

\acknowledgments
We thank Tabea Herrmann and Roland Ketzmerick for useful discussions.
This work was funded by the Deutsche Forschungsgemeinschaft
(DFG, German Research Foundation) --  497038782 (MK, AB) and by the European
Union's Horizon Europe program under the Marie Sk{\l}odowska Curie Action
SemiLiom (Grant No.~101198880) (MK).

\begin{appendix}
	\appendix

\section{Calculations for recursive embedding}\label{app:recursive_emmbedding}

To obtain the result \eqref{eq:discrete-diff-eq-for-mu} for the average purity
we start from the Schmidt decomposition of an eigenstate $\ket{a(\epsilon)}$.
We denote the Schmidt values of an eigenstate $\ket{a(\epsilon)}$ by
$\lambda_{a,1}, \lambda_{a,2}, \ldots$, such that
\begin{equation}
	\ket{a(\epsilon)} =
		\sqrt{\lambda_{a, 1}}\ket{m_{a, 1},n_{a, 1}}
			+ \sqrt{\lambda_{a, 2}}\ket{m_{a, 2},n_{a, 2}} + \ldots,
\end{equation}
where $\{\ket{m_i,n_j}\}$ is a product basis of the Hilbert space. For
$\epsilon \to \epsilon + \Delta \epsilon$ the new eigenstates are given by a
superposition of eigenstates at $\epsilon$, see
Eq.~\eqref{recursive_transition}. In order to identify a recursive
differential equation we impose some assumptions: First, we assume that for
all eigenstates $\ket{b(\epsilon)}$ and $\ket{b'(\epsilon)}$ contributing to
this superposition, we have $m_{b,i} \neq m_{b',j}$ and $n_{b,i} \neq
n_{b',j}$ for all $i,j$. This is true for $\epsilon = 0$ and plausible for
localized eigenstates and large Hilbert space dimensions. Then the Schmidt
decomposition of $\ket{a(\epsilon + \Delta \epsilon)}$ is given by
\begin{align}\label{eq:schmidt-decomposition-perturbative}
	&\biggr\{ \left( 1 - \Delta \epsilon^2 \sum_{b \neq a}
		\frac{|\matrixel{a(\epsilon)}{V}{b(\epsilon)}|^2}
			 {(\theta_{a(\epsilon)}-\theta_{b(\epsilon)})^2} \right)
			 \lambda_{a,i} \biggr\} \nonumber \\
	\cup~ &\biggr\{ \left( \Delta \epsilon^2
		\frac{|\matrixel{b(\epsilon)}{V}{a(\epsilon)}|^2}
		     {(\theta_{a(\epsilon)}-\theta_{b(\epsilon)})^2} \right)
			    \lambda_{b,i} \biggr\} , \nonumber \\
	\cup~ &\biggr\{ \left( \Delta \epsilon^2
		\frac{|\matrixel{b'(\epsilon)}{V}{a(\epsilon)}|^2}
		     {(\theta_{a(\epsilon)}-\theta_{b'(\epsilon)})^2} \right)
				\lambda_{b',i} \biggr \}, \nonumber \\
    &\hspace{2.5cm}\vdots
\end{align}
where the first line corresponds to a perturbative prefactor times the Schmidt
eigenvalues of $\ket{a(\epsilon)}$, the second to $\ket{b(\epsilon)}$ and so
forth. Additionally we assume the perturbative prefactors in
Eq.~\eqref{eq:schmidt-decomposition-perturbative} and the Schmidt values
$\lambda_{a, 1}, \lambda_{b,1},...$ to be independent as random variables. Then we arrive at
\begin{equation}\label{eq:recursive_differential_equation}
	\overline{\mu(\epsilon + \Delta \epsilon)} - \overline{\mu(\epsilon)}
	= -\left( 1 - v^2(\epsilon,\Delta \epsilon)
			   - v'^2(\epsilon, \Delta \epsilon)
	  \right) \overline{\mu(\epsilon)},
\end{equation}
where we denote
\begin{align}
	v^2(\epsilon,\Delta \epsilon) &=
	\overline{\left( 1 - \Delta \epsilon^2 \sum_{b \neq a}
		\frac{|\matrixel{a(\epsilon)}{V}{b(\epsilon)}|^2}
			 {(\theta_{a(\epsilon)}-\theta_{b(\epsilon)})^2} \right)^2} \label{eq:recursive_moments_1} \\
	v'^2(\epsilon,\Delta \epsilon) &=
	\overline{\sum_{b \neq a} \left(\Delta \epsilon^2
		\frac{|\matrixel{b(\epsilon)}{V}{a(\epsilon)}|^2}
			{(\theta_{a(\epsilon)}-\theta_{b(\epsilon)})^2}
				\right)^2}.
	\label{eq:recursive_moments_2}
\end{align}
We compute $v^2(\epsilon, \Delta \epsilon)$ and $v'^2(\epsilon, \Delta
\epsilon)$ in leading order of $\epsilon$. Then the eigenstates take the form
$\ket{a(\epsilon)} = \sqrt{\lambda_{a1}} \ket{jk, q_{\uA,a}} +
\sqrt{\lambda_{a2}} \ket{j'k', q'_{\uA,a}}$. Due to the form of the coupling
$V$ the corresponding values of $q_{\uA,a}$ fulfill $q_{\uA,a} = q_{\uA,a}'
\pm 1$. To make the following discussion explicit we consider states with
largest Schmidt value located in a block with fixed value $q_{\uA, a}$.
Without loss of generality, we assume that we have $q_{\uA,a} = q_{\uA,a}' -
1$. Furthermore we approximate the dimensions and variances,
Eq.~\eqref{eq:coupling_variance}, of all symmetry sectors appearing in the following calculation to be equal. Moreover, we assume that the
probabilities for $q_{\uA,b} = q_{\uA,b}' + 1$ and of $q_{\uA,b} = q_{\uA,b}'
- 1$ are both $1/2$. Even though these assumptions are not exact for CQTE
ensembles adapted for e.g.~Bose-Hubbard models, they remain a suitable
approximation for large system sizes. We abbreviate the sum over one block
with value $q_\uA'$ by
\begin{align}
	\sigma(\epsilon; q_\uA', \alpha) = \Delta \epsilon^{2\alpha}
		\sum_{\substack{b \neq a \\ q_{\uA, b} = q_\uA'}}
		\left[
			\frac{|\matrixel{a(\epsilon)}{V}{b(\epsilon)}|^2}
				 {(\theta_{a(\epsilon)}-\theta_{b(\epsilon)})^2}
		\right]^\alpha,
\end{align}
where we introduce the parameter $\alpha$ to describe arbitrary moments.
Equations ~\eqref{eq:recursive_moments_1} and ~\eqref{eq:recursive_moments_2}
can then be rewritten as in Ref.~\cite{TomLakSriBae2018},
\begin{align}
	v^2(\epsilon, \Delta \epsilon) &=
			1 - \sum_{i=-2}^3 2~\sigma(\epsilon; q_{\uA,a} + i, 1)
					\nonumber \\
		&\qquad + \sum_{i=-2}^3 \sigma(\epsilon; q_{\uA,a} + i, 2) \\
	v'^2(\epsilon, \Delta \epsilon) &=
			\sum_{i=-2}^3 \sigma(\epsilon; q_{\uA,a} + i, 2).
\end{align}
We investigate the contributions from different symmetry sectors $i$ one after
another, starting with $i=0$. We proceed in a similar manner to
Ref.~\cite{TomLakSriBae2018} by first finding the mean squared matrix element.
Them, the normalized distribution of the matrix elements which are
not identically zero due to the coupling is approximated by a Porter Thomas
distribution (PT) in the COE case and in the an exponential distribution (exp)in the CUE case.
We assume that the level statistics of the perturbed eigenstates are still
described by Poisson statistics, as the first order correction of the
eigenphases $\matrixel{jk}{V}{jk}$ vanishes. Then we use the mean level
spacing to take the ensemble average.
We start with $\sigma(\epsilon; q_{\uA,a},\alpha)$. The matrix element in terms of the
unperturbed eigenvectors is given by
\begin{align}
	&|\matrixel{a(\epsilon)}{V}{b(\epsilon)}|^2 = \nonumber \\
	&\qquad \left| \sqrt{\lambda_{a,1}\lambda_{b,2}}
		\matrixel{jk,q_{\uA,a}}{V}{m'n',q_{\uA,a}\pm 1} \right. \nonumber \\
    &\qquad\quad + \left. \sqrt{\lambda_{a,2}\lambda_{b,1}}
		\matrixel{j'k', q_{\uA,a}-1}{V}{mn, q_{\uA,a}}| \right|^2 \\
	&\quad = \lambda_{a,1}\lambda_{b,2}|
		\matrixel{jk,q_{\uA,a}}{V}{m'n',q_{\uA,a}\pm 1}|^2 \nonumber \\
	&\qquad\quad + \lambda_{a,2}\lambda_{b,1}
		|\matrixel{j'k', q_{\uA,a}-1}{V}{mn, q_{\uA,a}}|^2 \nonumber \\
	&\qquad\quad + 2\sqrt{\lambda_{a,1}\lambda_{b,2}
			\lambda_{a,2}\lambda_{b,1}} \nonumber \\
    &\qquad\qquad \times \Re
		\matrixel{j'k', q_{\uA,a}-1}{V}{mn, q_{\uA,a}} \cross \nonumber \\
	& \hspace{3.2cm} \cross \matrixel{m'n', q_{\uA,a}\pm 1 }{V}{jk, q_{\uA,a}}.
	\label{eq:new_matrix_element}
\end{align}
We neglect correlations between the matrix elements and the Schmidt values by
replacing the latter with the ensemble averages calculated in
Sec.~\ref{sec:entanglement_transition}. Hence, in leading order of $\epsilon$
we have
\begin{equation}
	\lambda_{a,1}\lambda_{b,2} \rightarrow
		\langle \lambda_1 \rangle
		\langle \lambda_2 \rangle
	\approx \langle \lambda_2 \rangle.
\end{equation}
By using that the last term in \eqref{eq:new_matrix_element} vanishes in the
ensemble average we obtain the approximated mean value of the squared matrix
element as $2\langle \lambda_2 \rangle \beta \sigma_{q_{\uA,a}}^2$. After
regularizing the sum $\sigma(\epsilon; q_\uA, \alpha)$, following
Ref.~\cite{TomLakSriBae2018}, we obtain the moments
\begin{equation}
	\sigma(\epsilon; q_\uA, \alpha) =
			\frac{\Delta\sqrt{\Lambda_{q_\uA}}}{2}
				\sqrt{2\langle \lambda_2 \rangle}
					c^{\text{PT/exp}}_2(\alpha).
\end{equation}
Here $\Delta \sqrt{\Lambda_{q_\uA}}$ is defined as a function of $\Delta
\epsilon$ similar to Eq.~\eqref{eq:gen_block_transition_parameter}, and the
constant $c^{\text{PT/exp}}_2(\alpha)$ is given by
\begin{align}
	c^{\text{PT/exp}}_2(\alpha) &=
			\int_{0}^{\infty} \ud s~\int_{0}^{\infty} \ud w~
				\frac{\varrho_{\text{PT/exp}}(\omega)}{2^{\alpha-1}} \times
					\nonumber \\
	&\qquad\qquad\qquad\quad \times  \left( 1 - \frac{s}{\sqrt{s^2 + 4 \omega}}
				\right)^\alpha.
\end{align}
The dependence of $\sigma(q_\uA, \alpha)$ on $\epsilon$ originates from the
factor $\sqrt{\lambda_2}$. Following the same procedure we obtain
\begin{align}
	\sigma(\epsilon; q_\uA + 2, \alpha)
		&= \frac{\Delta \sqrt{\Lambda_{q_\uA}}}{4}
				\sqrt{2\langle \lambda_2 \rangle}
					c^{\text{PT/exp}}_2(\alpha) \nonumber \\
		&\qquad +\frac{\Delta \sqrt{\Lambda_{q_\uA}}}{4}
				\sqrt{\langle \lambda_2 \rangle}
					c^{\text{PT/exp}}_2(\alpha) \\
	\sigma(\epsilon; q_\uA + 3, \alpha)
		&= \frac{\Delta\sqrt{\Lambda_{q_\uA}}}{4} \langle \lambda_2 \rangle
				c^{\text{PT/exp}}_2(\alpha) \\
	\sigma(\epsilon; q_\uA - 2, \alpha)
		&= \frac{\Delta \sqrt{\Lambda_{q_\uA}}}{4}
			\sqrt{\langle \lambda_2 \rangle}
				c^{\text{PT/exp}}_2(\alpha).
\end{align}
Note that the factor $1/4$ appears because half of the matrix elements
(where $q_{\uA,b} = q_{\uA,b}' + 1$ ($q_{\uA,b} = q_{\uA,b}' - 1$)) are
identically zero due to the structure of the coupling. The moments of the remaining sums are
given by
\begin{align}
	\sigma(\epsilon; q_\uA + 1,\alpha)
		&= \frac{1}{2}\Delta \sqrt{\Lambda_{q_\uA}}
			\sqrt{\langle \lambda_1 \rangle ^2 + \langle \lambda_2 \rangle^2}
				c^{\text{PT/exp}}_2(\alpha) \\
		&\approx \frac{1}{2} \Delta \sqrt{\Lambda_{q_\uA}}
			\langle \lambda_1 \rangle c^{\text{PT/exp}}_2(\alpha) \\
	\sigma(\epsilon; q_\uA-1,\alpha)
		&= \frac{1}{2} \Delta \sqrt{\Lambda_{q_\uA}}
			\langle \lambda_1 \rangle c^{\text{PT/exp}}_2(\alpha).
\end{align}
These results can now be combined to yield $v^2(\epsilon, \Delta \epsilon)$
and $v'^2(\epsilon, \Delta \epsilon)$ and the differential equation
\eqref{eq:recursive_differential_equation}.
We arrive at the solution of Eq.~\eqref{eq:recursive_differential_equation}
\begin{align}
	\overline{\mu_{q_\uA}}(\Lambda_{q_\uA}) &\approx \ue^{-s(\Lambda_{q_\uA})},
\end{align}
where for the COE
\begin{align}
s(\Lambda_{q_\uA}) &=
			\sqrt{2\pi} \sqrt{\Lambda_{q_\uA}}
		  + (2\pi)^{1/4}\frac{3\sqrt{2}+2}{3}\Lambda_{q_\uA}^{3/4}
		  - \frac{3}{2} \Lambda_{q_\uA},
\end{align}
and for the CUE
\begin{align}
	s(\Lambda_{q_\uA}) &=
			\frac{\pi^{3/2}}{2} \sqrt{\Lambda_{q_\uA}}
		  + \pi^{7/4}\frac{3\sqrt{2}+2}{12}\Lambda_{q_\uA}^{3/4}
	      - \frac{3}{16} \pi^2 \Lambda_{q_\uA}.
\end{align}

\end{appendix}


\begin{thebibliography}{10}
\newcommand{\enquote}[1]{``#1''}
\providecommand{\url}[1]{\texttt{#1}}
\providecommand{\urlprefix}{URL }
\providecommand{\eprint}[2][]{\url{#2}}

\bibitem{AmiFazOstVed2008}
L.~Amico, R.~Fazio, A.~Osterloh, and V.~Vedral, \emph{Entanglement in many-body
  systems}, Rev.~Mod.~Phys. \textbf{80}, 517 (2008).

\bibitem{NanHus2015}
R.~Nandkishore and D.~A. Huse, \emph{Many-body localization and thermalization
  in quantum statistical mechanics}, Annu.~Rev.~Condens.~Matter Phys.
  \textbf{6}, 15 (2015).

\bibitem{AbaPap2017}
D.~A. Abanin and Z.~Papi{\'c}, \emph{Recent progress in many-body
  localization}, Ann.~Phys.~(Berlin) \textbf{529}, 1700169 (2017).

\bibitem{AbaAltBloSer2019}
D.~A. Abanin, E.~Altman, I.~Bloch, and M.~Serbyn, \emph{Colloquium: Many-body
  localization, thermalization, and entanglement}, Rev.~Mod.~Phys. \textbf{91},
  021001 (2019).

\bibitem{GolSel2018}
M.~Goldstein and E.~Sela, \emph{Symmetry-resolved entanglement in many-body
  systems}, Phys.~Rev.~Lett. \textbf{120}, 200602 (2018).

\bibitem{FelGol2019}
N.~Feldman and M.~Goldstein, \emph{Dynamics of charge-resolved entanglement
  after a local quench}, Phys.~Rev.~B \textbf{100}, 235146 (2019).

\bibitem{BonRugCal2019}
R.~Bonsignori, P.~Ruggiero, and P.~Calabrese, \emph{Symmetry resolved
  entanglement in free {Fermi}onic systems}, J.~Phys.~A \textbf{52}, 475302
  (2019).

\bibitem{CapRugCal2020}
L.~Capizzi, P.~Ruggiero, and P.~Calabrese, \emph{Symmetry resolved entanglement
  entropy of excited states in a {CFT}}, J.~Stat.~Mech. \textbf{2020}, 073101
  (2020).

\bibitem{MurDubCal2024}
S.~Murciano, J.~Dubail, and P.~Calabrese, \emph{More on symmetry resolved
  operator entanglement}, J.~Phys.~A \textbf{57}, 145002 (2024).

\bibitem{KieUnaFleSir2020}
M.~{Kiefer-Emmanouilidis}, R.~Unanyan, M.~Fleischhauer, and J.~Sirker,
  \emph{Evidence for unbounded growth of the number entropy in many-body
  localized phases}, Phys.~Rev.~Lett. \textbf{124}, 243601 (2020).

\bibitem{KieUnaFleSir2021}
M.~{Kiefer-Emmanouilidis}, R.~Unanyan, M.~Fleischhauer, and J.~Sirker,
  \emph{Slow delocalization of particles in many-body localized phases},
  Phys.~Rev.~B \textbf{103}, 024203 (2021).

\bibitem{KieUnaSirFle2020}
M.~{Kiefer-Emmanouilidis}, R.~Unanyan, J.~Sirker, and M.~Fleischhauer,
  \emph{Bounds on the entanglement entropy by the number entropy in
  non-interacting {Fermi}onic systems}, SciPost Phys. \textbf{8}, 083 (2020).

\bibitem{LukRisSchTaiKauChoKheLeoGre2019}
A.~Lukin, M.~Rispoli, R.~Schittko, M.~E. Tai, A.~M. Kaufman, S.~Choi,
  V.~Khemani, J.~L{\'e}onard, and M.~Greiner, \emph{Probing entanglement in a
  many-body -- localized system}, Science \textbf{364}, 256 (2019).

\bibitem{NevCarVitKokElbDalCalZolVerKueKra2021}
A.~Neven, J.~Carrasco, V.~Vitale, C.~Kokail, A.~Elben, M.~Dalmonte,
  P.~Calabrese, P.~Zoller, B.~Vermersch, R.~Kueng, and B.~Kraus,
  \emph{Symmetry-resolved entanglement detection using partial transpose
  moments}, npj Quantum Information \textbf{7}, 152 (2021).

\bibitem{VitElbKueNevCarKraZolCalVerDal2022}
V.~Vitale, A.~Elben, R.~Kueng, A.~Neven, J.~Carrasco, B.~Kraus, P.~Zoller,
  P.~Calabrese, B.~Vermersch, and M.~Dalmonte, \emph{Symmetry-resolved
  dynamical purification in synthetic quantum matter}, SciPost Phys.
  \textbf{12}, 106 (2022).

\bibitem{SriTomLakKetBae2016}
S.~C.~L. Srivastava, S.~Tomsovic, A.~Lakshminarayan, R.~Ketzmerick, and
  A.~B\"acker, \emph{Universal scaling of spectral fluctuation transitions for
  interacting chaotic systems}, Phys.~Rev.~Lett. \textbf{116}, 054101 (2016).

\bibitem{TomLakSriBae2018}
S.~Tomsovic, A.~Lakshminarayan, S.~C.~L. Srivastava, and A.~B\"acker,
  \emph{Eigenstate entanglement between quantum chaotic subsystems: Universal
  transitions and power laws in the entanglement spectrum}, Phys.~Rev.~E
  \textbf{98}, 032209 (2018).

\bibitem{LakSriKetBaeTom2016}
A.~Lakshminarayan, S.~C.~L. Srivastava, R.~Ketzmerick, A.~B{\"a}cker, and
  S.~Tomsovic, \emph{Entanglement and localization transitions in eigenstates
  of interacting chaotic systems}, Phys.~Rev.~E \textbf{94}, 010205(R) (2016).

\bibitem{BerBalTabVor1979}
M.~V. Berry, N.~L. Balazs, M.~Tabor, and A.~Voros, \emph{Quantum maps},
  Ann.~Phys.~(N.Y.) \textbf{122}, 26 (1979).

\bibitem{ChaLucCha2018a}
A.~Chan, A.~De~Luca, and J.~T. Chalker, \emph{Spectral statistics in spatially
  extended chaotic quantum many-body systems}, Phys.~Rev.~Lett. \textbf{121},
  060601 (2018).

\bibitem{Pro2000}
T.~Prosen, \emph{Exact time-correlation functions of quantum {Ising} chain in a
  kicking transversal magnetic field}, Prog.~Theor.~Phys.~Suppl. \textbf{139},
  191 (2000).

\bibitem{PulLakSriBaeTom2020}
J.~J. Pulikkottil, A.~Lakshminarayan, S.~C.~L. Srivastava, A.~B{\"a}cker, and
  S.~Tomsovic, \emph{Entanglement production by interaction quenches of quantum
  chaotic subsystems}, Phys.~Rev.~E \textbf{101}, 032212 (2020).

\bibitem{HerKieFriBae2020}
T.~Herrmann, M.~F.~I. Kieler, F.~Fritzsch, and A.~B{\"a}cker,
  \emph{Entanglement in coupled kicked tops with chaotic dynamics},
  Phys.~Rev.~E \textbf{101}, 022221 (2020).

\bibitem{PulLakSriKieBaeTom2023}
J.~J. Pulikkottil, A.~Lakshminarayan, S.~C.~L. Srivastava, M.~F.~I. Kieler,
  A.~B{\"a}cker, and S.~Tomsovic, \emph{Quantum coherence controls the nature
  of equilibration and thermalization in coupled chaotic systems}, Phys.~Rev.~E
  \textbf{107}, 024124 (2023).

\bibitem{FriKie2024}
F.~Fritzsch and M.~F.~I. Kieler, \emph{Universal spectral correlations in
  interacting chaotic few-body quantum systems}, Phys.~Rev.~E \textbf{109},
  014202 (2024).

\bibitem{YosGarCha2025}
T.~Yoshimura, S.~J. Garratt, and J.~T. Chalker, \emph{Operator dynamics in
  {Floquet} many-body systems}, Phys.~Rev.~B \textbf{111}, 094316 (2025).

\bibitem{AltTelMic2024}
A.~Altland, J.~{Telles de Miranda}, and T.~Micklitz, \emph{Tensor product
  random matrix theory}, Phys.~Rev.~Res. \textbf{6}, L042029 (2024).

\bibitem{AltKimMic2025:p}
A.~Altland, K.~W. Kim, and T.~Micklitz, \emph{Path integral approach to quantum
  thermalization}, arXiv:2509.06028 [cond-mat.mes-hall]  (2025).

\bibitem{KieFriBae2026:p}
M.~F.~I. Kieler, F.~Fritzsch, and A.~B{\"a}cker, \emph{Semiclassical foundation
  of universality in chaotic quantum circuits}, arXiv:2605.27052 [quant-ph]
  (2026).

\bibitem{RosPor1960}
N.~Rosenzweig and C.~E. Porter, \emph{"{Repulsion} of energy levels" in complex
  atomic spectra}, Phys. Rev. \textbf{120}, 1698 (1960).

\bibitem{KraKhaCueAmi2015}
V.~E. Kravtsov, I.~M. Khaymovich, E.~Cuevas, and M.~Amini, \emph{A random
  matrix model with localization and ergodic transitions}, New J. Phys.
  \textbf{17}, 122002 (2015).

\bibitem{BuiBar2022}
W.~Buijsman and Y.~Bar~Lev, \emph{Circular {Rosenzweig}-{Porter} random matrix
  ensemble}, SciPost Phys. \textbf{12}, 082 (2022).

\bibitem{Efe1983}
K.~Efetov, \emph{Supersymmetry and theory of disordered metals}, Adv.~Phys.
  \textbf{32}, 53 (1983).

\bibitem{FyoMir1994}
Y.~V. Fyodorov and A.~D. Mirlin, \emph{Statistical properties of eigenfunctions
  of random quasi 1d one-particle {H}amiltonians}, Int.~J.~Mod.~Phys.~B
  \textbf{08}, 3795 (1994).

\bibitem{AltZir1996}
A.~Altland and M.~R. Zirnbauer, \emph{Field theory of the quantum kicked
  rotor}, Phys.~Rev.~Lett. \textbf{77}, 4536 (1996).

\bibitem{BarHerDel2018}
H.~Barghathi, C.~M. Herdman, and A.~Del~Maestro, \emph{R{\'e}nyi generalization
  of the accessible entanglement entropy}, Phys.~Rev.~Lett. \textbf{121},
  150501 (2018).

\bibitem{BarCasDel2019}
H.~Barghathi, E.~{Casiano-Diaz}, and A.~Del~Maestro, \emph{Operationally
  accessible entanglement of one-dimensional spinless {Fermi}ons}, Phys.~Rev.~A
  \textbf{100}, 022324 (2019).

\bibitem{MurCalPir2022}
S.~Murciano, P.~Calabrese, and L.~Piroli, \emph{Symmetry-resolved page curves},
  Phys.~Rev.~D \textbf{106}, 046015 (2022).

\bibitem{BiaDon2019}
E.~Bianchi and P.~Don{\`a}, \emph{Typical entanglement entropy in the presence
  of a center: Page curve and its variance}, Phys.~Rev.~D \textbf{100}, 105010
  (2019).

\bibitem{TkoSmaKusZeiZyc2012}
T.~Tkocz, M.~Smaczy\'nski, M.~Ku\'s, O.~Zeitouni, and K.~\.Zyczkowski,
  \emph{Tensor products of random unitary matrices}, Random Matrices: Theory
  Appl. \textbf{1}, 1250009 (2012).

\bibitem{StrGraKor2008}
M.~P. Strzys, E.~M. Graefe, and H.~J. Korsch, \emph{Kicked {Bose}--{Hubbard}
  systems and kicked tops --- destruction and stimulation of tunneling}, New J.
  Phys. \textbf{10}, 013024 (2008).

\bibitem{FavFazRus2020}
M.~Fava, R.~Fazio, and A.~Russomanno, \emph{Many-body dynamical localization in
  the kicked {Bose}-{Hubbard} chain}, Phys.~Rev.~B \textbf{101}, 064302 (2020).

\bibitem{Kol2016}
A.~R. Kolovsky, \emph{Bose--{Hubbard} {H}amiltonian: {Quantum} chaos approach},
  Int.~J.~Mod.~Phys.~B \textbf{30}, 1630009 (2016).

\bibitem{DubMue2016}
R.~Dubertrand and S.~M{\"u}ller, \emph{Spectral statistics of chaotic many-body
  systems}, New J. Phys. \textbf{18}, 033009 (2016).

\bibitem{RicUrbTom2022}
K.~Richter, J.~D. Urbina, and S.~Tomsovic, \emph{Semiclassical roots of
  universality in many-body quantum chaos}, J.~Phys.~A \textbf{55}, 453001
  (2022).

\bibitem{BiaDonKum2024}
E.~Bianchi, P.~Dona, and R.~Kumar, \emph{Non-{Abelian} symmetry-resolved
  entanglement entropy}, SciPost Phys. \textbf{17}, 127 (2024).

\end{thebibliography}
\end{document}